\documentclass[prl, twocolumn,showpacs,nofootinbibfloatfix,amsmath,amsfonts,amssymb]{revtex4-1}%
\usepackage{amsmath,amsfonts,amssymb,color}
\usepackage{graphicx}
\usepackage{dcolumn}
\usepackage{bm}

\begin{document}
\title{$Z_4$ parafermion $\pm \pi/2$ modes in interacting periodically driven superconducting chain}
\author{Raditya Weda Bomantara}
\email{Raditya.Bomantara@sydney.edu.au}
\affiliation{%
	Centre for Engineered Quantum Systems, School of Physics, University of Sydney, Sydney, New South Wales 2006, Australia
}
\date{\today}


\vspace{2cm}

\begin{abstract}
	We theoretically report the emergence of $Z_4$ parafermion edge modes in a periodically driven spinful superconducting chain with modest fermionic Hubbard interaction. These parafermion edge modes represent $\pm \pi/(2T)$ quasienergy excitations ($T$ being the driving period), which have no static counterpart and arise from the interplay between interaction effect and periodic driving. At special parameter values, these exotic quasiparticles can be analytically and exactly derived. Strong numerical evidence of their robustness against variations in parameter values and spatial disorder is further presented. Our proposal offers a route toward realizing parafermions without fractional quantum Hall systems or complicated interactions.
	
\end{abstract}

\maketitle

{\it Introduction.} Topological phases of matter have been the subject of extensive theoretical \cite{Kit,Kane2005, Bernevig2006,FuKane2007,Fu2007,Moore2007,Schnyder2008,Kit2,Wan2011,DLoss,Bomantara2017,Tuloup2020,HTI1,HTI2,LU} and experimental \cite{Konig2007,Hsieh2008,burger2013,Xu2015,Lv2015} studies since the last two decades. Among their remarkable features is their insensitivity to local perturbations, which is expected to find a promising application in the development of fault-tolerant quantum computers \cite{Nayak2008,QEC5,tqc2}. For this specific purpose, topological phases exhibiting anyonic quasiparticle excitations are especially sought after due to their ability to encode quantum information nonlocally. Majorana fermions are the well-studied examples of such quasiparticle excitations that are expected to exist at the edges of superconductivity proximitized nanowires \cite{sc1,sc2}. Over the years, various schemes for braiding pairs of Majorana fermions have been theoretically devised \cite{braid1,braid2,braid3,braid4,braid5,braid6,braid7,braid8,braid9}, and their integration into a qubit architecture has been proposed \cite{mr1,mr2,mr3,mr4}. On the experimental side, although their quantum computing capabilities have yet to be confirmed, some signatures of Majorana fermions have been successfully observed \cite{Majsig1,Majsig2,Majsig3,Majsig4,Majsig5}.

While Majorana fermions are particularly attractive due to their experimental feasibility, they do not support the full set of topologically protected gates for quantum universality \cite{nonuni}. In particular, braiding of Majorana fermions will only yield either a Pauli $Z$, Pauli $X$, or a controlled-Pauli gate, which together forms the so-called Clifford gate set \cite{cliff}. $Z_n$ parafermions \cite{par1,par2} are more exotic quasiparticle excitations generalizing Majorana fermions ($Z_2$ parafermions) which, while generally still do not enable quantum universality, support a richer set of topologically protected quantum gates. As compared with Majorana-based qubits, parafermion-based qubits thus benefit from requiring lower space overhead for the implementation of quantum error correction and magic state distillation, both of which are extremely important components of future quantum computers.   

Unlike Majorana fermions, studies of parafermions are relatively scarce \cite{parstudies,parstudies2,parstudies3,parstudies4,parstudies5} and have still left many open questions. Theoretically, due to the necessity for considering specific strongly interacting systems in their studies, band structure analysis typically employed for characterizing ``free" fermion systems, including those supporting Majorana fermions, is no longer feasible. Sophisticated methods and limited numerical calculations have thus been utilized in these existing studies, obscuring the accessibility of such parafermionic systems. Experimentally, well-known proposals for realizing parafermions \cite{parprop,parprop2,parprop3} require access to the elusive fractional quantum Hall systems. While Refs.~\cite{parMaj,parMaj2,parMaj3,parMaj4} suggest that $Z_4$ parafermionic zero modes can in principle exist without the aid of fractional quantum Hall systems, they require the presence of intricate interaction effects that are challenging to realize in experiments.            

Motivated by the above challenges, we propose a potentially simpler platform for realizing $Z_4$ parafermions. Our proposal only requires a spinful $p$-wave superconductor, fermionic Hubbard interaction, tunable magnetic field, and appropriate periodic driving, all of which can be achieved with current technology. In particular, a $p$-wave superconductor, which can be obtained by proximitizing a nanowire or an edge of topological insulator with an $s$-wave superconductor, has been successfully achieved in various experiments \cite{Majsig1,Majsig2,Majsig3,Majsig4,Majsig5} for studies of Majorana fermions. Apart from its experimental feasibility, our proposal demonstrates the ability of periodic driving and interaction effect to generate unique topological features, thus advancing the area of Floquet topological phases \cite{Flor6,Flor7,Flor8,Flor11,Flor12,Rud,Flor15,Flor18,Flor22,Flor23,Flor24} (see also \cite{Frev1,Frev2} for comprehensive reviews) to interacting setting.

{\it Model.} Our construction is inspired by the following two realizations: 1) a single Kitaev chain \cite{Kit} in the topologically nontrivial regime can be understood as a single nonlocal qubit (formed by its two Majorana modes at the two ends) and 2) one can devise a unitary operation that realizes a particular two-qubit quantum gate $G$, which consists of a Pauli-X gate followed by a CNOT gate and satisfies a $Z_4$ symmetry, i.e., $G^4=1$ but $G^k\neq 1$ for $k=1,2,3$. In this case, a system of two topologically nontrivial Kitaev chains, which is achieved, e.g., with a spinful $p$-wave superconductor, supports totally two nonlocal qubits. By further devising a periodic driving scheme such that the system's one-period time evolution (Floquet) operator simulates the gate $G$ acting on the two qubits, a pair of $Z_4$ parafermionic modes is then expected to emerge at the system's ends as a result of the entanglement between the two qubits. As further elaborated below, we find that such a Floquet operator can be realized through the following five-time-step Hamiltonian, i.e., 
\begin{eqnarray}
H(t) &=& H_\ell \; \text{ for } \frac{\ell}{5}<\frac{t}{T}<\frac{\ell+1}{5}\;, \nonumber \\
H_1 &=& \sum_{j=1}^{\frac{N}{2}}\sum_{s=\pm 1}(H_{2j-1,s}^{(\rm onsite)}-H_{2j,s}^{(\rm Zeeman)}) \;, \nonumber \\
H_2 &=&\sum_{j=1}^{\frac{N}{2}-1}(H_{2j-1,1}^{(\rm Kitaev)}+H_{2j,-1}^{(\rm Kitaev)}) ,\; H_3=\sum_{j=1}^N H_j^{\rm (int)} \;, \nonumber \\
H_4&=&\sum_{j=1}^{\frac{N}{2}-1}(H_{2j-1,-1}^{(\rm Kitaev)}+H_{2j,1}^{(\rm Kitaev)}) \;, \nonumber \\
H_5&=&\sum_{j=1}^{\frac{N}{2}}\sum_{s=\pm 1}(H_{2j,s}^{(\rm onsite)}+H_{2j-1,s}^{(\rm Zeeman)}) \;, \label{sysraw} 
\end{eqnarray}
over the course of some period $T$. Specifically, 
\begin{eqnarray}
H_{j,s}^{\rm (onsite)} &=& s\mu_{j,s}\hat{n}_{j,s} \;,\; H_{j,s}^{\rm (Zeeman)} =J_j^{(1)} c_{j,s}^\dagger c_{j,-s} \;, \nonumber \\
H_{j,s}^{\rm (Kitaev)} &=& -J_{j,s}^{(2)} c_{j+1,s}^\dagger c_{j,s} +\Delta_{j,s} c_{j+1,s}^\dagger c_{j,s}^\dagger +h.c. \;, \nonumber \\
H_{j,s}^{\rm (int)} &=& U_j \hat{n}_{j,1} \hat{n}_{j,-1} \;,
\end{eqnarray}  
where $\mu_{j,s}$, $J_{j,s}^{(1)}$, $J_{j,s}^{(2)}$ and $\Delta_{j,s}$ are respectively the chemical potentials, Zeeman field, hopping amplitude, and $p$-wave pairing, $N$ is the system size, $U_{j}$ is the fermionic Hubbard interaction strength, $c_{j,s}$ is the fermion operator associated with spin $s$ at site $j$, and $\hat{n}_{j,s}=c_{j,s}^\dagger c_{j,s}$. 

While such a driving protocol may seem artificial, we note that a similar quenched driving scheme has been experimentally achieved in several different contexts \cite{DTCexp1,DTCexp2,Zhu}. As such, quickly switching on/off system parameters simulating a square pulse is not unrealistic in practice. It is further noted that in the cold atoms proposal of Kitaev chain \cite{FMF1,braid3}, all system parameters introduced above should in principle be tunable to allow for periodic modulation. In the proximitized nanowire platform, a chain of Cooper pair boxes \cite{braid2,CPB} in the spirit of \cite{braid7} can in principle be utilized to introduce tunable parameters.

{\it Noninteracting limit.} To develop insight into the above model, we first consider the noninteracting limit, i.e., $U_j=0$, in which the system's full quasienergy ($\varepsilon$) excitation spectrum can be obtained by diagonalizing the Bogolioubov-de-Genes (BdG) unitary $\mathcal{U}_T = \mathcal{U}_5\times \mathcal{U}_4 \times \mathcal{U}_3 \times \mathcal{U}_2\times \mathcal{U}_1$, the eigenvalues of which take the form $e^{-\mathrm{i}\varepsilon T}$. Here, $\mathcal{U}_T$ is related to the actual second-quantized Floquet operator via $\tilde{U}_T=\frac{1}{2} \Psi^\dagger \mathcal{U}_T \Psi$, where $\Psi=\bigotimes_{j=1}^{\frac{N}{2}} \bigotimes_{s=\pm 1}^{\frac{N}{2}} \left(c_{2j-1,s}, c_{2j,s}, c_{2j-1,s}^\dagger, c_{2j,s}^\dagger \right)^T$. By further assuming the absence of spatial disorder, i.e., $\mu_{j,s}=\mu$, $J_{j}^{(1)}=J^{(1)}$, $J_{j,s}^{(2)}=\Delta_{j,s}=J^{(2)}$, we can explicitly write
\begin{eqnarray}
\mathcal{U}_1 &=& e^{-\mathrm{i} \sum_{j=1}^{\frac{N}{2}} \left( \frac{\mu T}{10} \sigma_z (1+\tau_z) -\frac{J^{(1)}T}{10} \sigma_x (1-\tau_z)\right) \eta_z|j\rangle \langle j| } \;,\nonumber \\
\mathcal{U}_4\times \mathcal{U}_3\times \mathcal{U}_2 &=&  e^{-\mathrm{i} \sum_{j=1}^{\frac{N}{2}-1} \frac{J^{(2)} T}{10} \left( \tau_x +\mathrm{i} \tau_y \right)\left( -\eta_z +\mathrm{i} \eta_y \right) |j+1\rangle \langle j |+h.c.} \nonumber \\
&& \times e^{-\mathrm{i} \sum_{j=1}^{\frac{N}{2}-1} \frac{J^{(2)} T}{5} \left( -\tau_x \eta_z + \tau_y \eta_y \right) |j\rangle \langle j | } \;, \nonumber \\
\mathcal{U}_5 &=&  e^{-\mathrm{i} \sum_{j=1}^{\frac{N}{2}} \left( \frac{\mu T}{10} \sigma_z (1-\tau_z) +\frac{J^{(1)} T}{10} \sigma_x (1+\tau_z)\right) \eta_z|j\rangle \langle j| } \;, \nonumber \\ \label{Eq:Nonint}
\end{eqnarray}
where $|j\rangle$ represents the $j$th site, $\sigma$, $\tau$, and $\eta$ are respectively Pauli matrices acting on the spin, sublattice, and particle-hole degrees of freedom. There, the simplification of $\mathcal{U}_4\times \mathcal{U}_3\times \mathcal{U}_2$ was obtained after utilizing $\mathcal{U}_3=1$ and $[H_{2j-1,s}^{\rm (Kitaev)},H_{2j,s}^{\rm (Kitaev)}]=0$ if $J_{j,s}^{(2)}=\Delta_{j,s}$.

Under periodic boundary conditions, Eq.~(\ref{Eq:Nonint}) can further be block-diagonalized into sectors of conserved quasi-momentum $k$, whose asssociated eight-bands momentum space Floquet operator $u_T(k)$ is easily obtained from Eq.~(\ref{Eq:Nonint}) by replacing $|j\rangle \langle j|\rightarrow |k\rangle \langle k| $ and $|j+1\rangle \langle j|\rightarrow e^{\mathrm{i} k} |k\rangle \langle k| $. It can then be easily verified that the system only respects the particle-hole symmetry $\mathcal{P} u_T(k) \mathcal{P}^{-1} =u_T(-k)$ with respect to $\mathcal{P}=\mathcal{K}$ ($\mathcal{K}$ being the complex conjugation operator), thus placing it in the D class \cite{Flosym} within the Altland-Zirnbauer classification \cite{AZ}. Such a particle-hole symmetry in turn enables the formation of in-gap topological edge states, such as the Majorana zero or $\pi$ modes \cite{parstudies5,FMF1,FMF2,kk3,FMF3,FMF7,FMF8,FMF9,FMF10}. 

\begin{center} 
	\begin{figure}
		\includegraphics[scale=0.5]{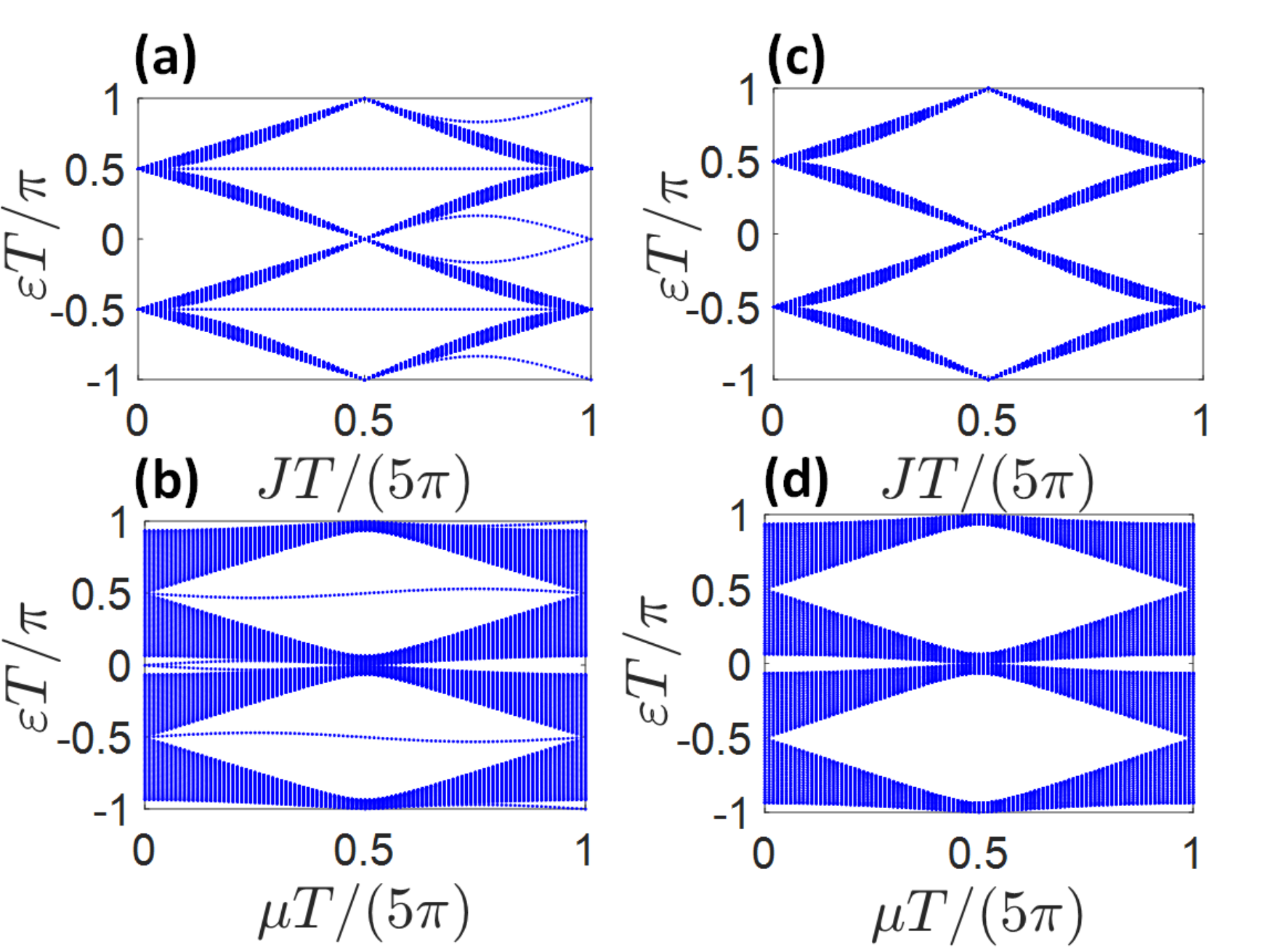}
		\caption{Quasienergy excitation spectrum of the system in the noninteracting limit under (a,b) open boundary conditions and (c,d) periodic boundary conditions at varying (a,c) $J^{(1)}=2J^{(2)}=J$ and (b,d) $\mu$. The other system parameters are fixed at (a,c) $\mu T/5=\pi/2$, (b,d) $J^{(1)}T/5=\pi/2-0.2$ and $J^{(2)}T/5=\pi/4$.}
		\label{fig:nonintspec}
	\end{figure}
\end{center}

As an important observation that we will exploit below, note that while Majorana modes can only exist as zero or $\pi/T$ quasienergy excitations, non-Majorana edge modes at other quasienergy excitations can in principle arise due to the absence of chiral symmetry. Indeed, as shown in Fig.~\ref{fig:nonintspec}, $\pm \pi/(2T)$ quasienergy edge modes can be clearly identified in the system under consideration. Analytically solving for these $\pm \pi/2$ modes at specific parameter values, as detailed in \cite{SI}, reveals that they represent ordinary fermions. In the following, we will show that the presence of interaction promotes these fermionic $\pm \pi/2$ modes into $Z_4$ parafermions.

{\it Parafermion $\pm \pi/2$ modes in the ideal case.} In the following, it will be useful to introduce Hermitian and mutually anticommuting Majorana operators $\gamma_{A,j,\pm 1}$ and $
\gamma_{B,j,\pm 1}$, such that $c_{j,\pm 1}^\dagger=\frac{\gamma_{B,j,\pm 1}+\mathrm{i} \gamma_{A,j,\pm 1}}{2}$. In the ideal case that $U_{j}=2\mu_{j,s}=2J_{j}^{(1)}=5\pi/T$ and $J_{j,s}^{(2)}=\Delta_{j,s}=J$, we may consider a rotated frame in which the system's Floquet operator can be written as \cite{SI}
\begin{eqnarray}
U_T &=& \overline{G} \times S \;, \nonumber \\
\overline{G} &=& e^{\sum_{j=1}^N \mathrm{i} \frac{\pi}{4} \gamma_{A,j,1}\gamma_{B,j,1}\gamma_{A,j,-1}\gamma_{B,j,-1} } \nonumber \\ 
&& \times e^{\sum_{j=1}^{N} \sum_{s=\pm 1} \frac{\pi}{4} \gamma_{B,j,s}\gamma_{A,j,-s} } \times e^{\sum_{j=1}^{N} \frac{\pi}{2} \gamma_{A,j,1}\gamma_{B,j,1} } \;, \nonumber \\
S &=& e^{\sum_{j=1}^{N-1} \mathrm{i} \frac{JT}{5} \gamma_{B,j,1}\gamma_{A,j+1,1}\gamma_{A,j,-1} \gamma_{B,j,-1}} \nonumber \\
&& \times e^{\sum_{j=1}^{N-1} \mathrm{i} \frac{JT}{5} \gamma_{B,j,-1}\gamma_{A,j+1,-1}\gamma_{A,j+1,1} \gamma_{B,j+1,1}} \;. \label{Floq}
\end{eqnarray}

{In Eq.~(\ref{Floq}), $S$ can be understood as the time evolution associated with two decoupled topologically nontrivial Kitaev chains (a more familiar form \cite{Kit} can be obtained by defining another set of Majorana operators $\tilde{\gamma}_{A,j,1}=\gamma_{A,j,1}$, $\tilde{\gamma}_{B,j,1}=\mathrm{i} \gamma_{B,j,1} \gamma_{A,j,-1}\gamma_{B,j,-1}$, $\tilde{\gamma}_{A,j,-1}=\mathrm{i} \gamma_{A,j,-1} \gamma_{A,j,1}\gamma_{B,j,1}$, and $\tilde{\gamma}_{B,j,-1}=\gamma_{B,j,-1}$). It supports two left-(right-)localized Majorana modes as $\gamma_1^{(L)}=\gamma_{A,1,1}$ ($\gamma_1^{(R)}=\gamma_{B,N,-1}$)  and $\gamma_2^{(L)}=\mathrm{i} \gamma_{A,1,1} \gamma_{B,1,1} \gamma_{A,1,-1}$ ($\gamma_2^{(R)}=\mathrm{i} \gamma_{B,N,1} \gamma_{A,N,-1} \gamma_{B,N,-1}$), which satisfy $[S,\gamma_i^{(P)}]=0$ for $i=1,2$ and $P=L,R$. Two nonlocal qubits can in turn be defined via the logical Pauli operators $(\overline{X}_1,\overline{Z}_1)=(\mathrm{i}\gamma_1^{(L)}\gamma_1^{(R)}, \gamma_1^{(L)}\gamma_2^{(L)})$ and $(\overline{X}_2,\overline{Z}_2)=(\gamma_1^{(L)}\gamma_2^{(L)}\gamma_1^{(R)}\gamma_2^{(R)}, \gamma_2^{(L)})$.
	
	The unitary $\overline{G}$ is designed such that it acts on the two qubits encoded by $S$ as a Pauli-X (on the first qubit) followed by a CNOT gate, i.e., it maps $\overline{X}_1\rightarrow \overline{X}_1 \overline{X}_2$, $\overline{X}_2\rightarrow \overline{X}_2$, $\overline{Z}_1\rightarrow -\overline{Z}_1 $, and $\overline{Z}_2\rightarrow \overline{Z}_1\overline{Z}_2$. As a consequence, it mixes $\gamma_1^{(L)}$ and $\gamma_2^{(L)}$, each of which is no longer a Majorana mode of $U_T$. Indeed, repeatedly conjugating $\gamma_1^{(L)}$ with $U_T$ yields \cite{SI}}
\begin{eqnarray}
U_T^\dagger \gamma_1^{(L)} U_T &=& -\gamma_2^{(L)} \;, \nonumber \\
U_T^\dagger \left(-\gamma_2^{(L)}\right) U_T  &=& -\gamma_1^{(L)} \;.
\end{eqnarray}
It then immediately follows that 
\begin{equation}
\psi_{\pm \pi/2}^L= e^{\mathrm{i} \frac{\pi}{4}} \left(\gamma_1^{(L)}\pm \mathrm{i} \gamma_2^{(L)} \right)/\sqrt{2}    
\end{equation}
satisfy $U_T^\dagger \psi_{\pm \pi/2}^L U_T = \mp \mathrm{i} \psi_{\pm \pi/2}^L$, thus corresponding to $\pm \pi/2$ modes. Moreover, it is easily verified that $(\psi_{\pm \pi/2}^{L})^2=\mathrm{i}\gamma_{B,1,1}\gamma_{A,1,-1}$ and $(\psi_{\pm\pi/2}^{L})^4=1$, i.e., $\psi^{L}_{\pm \pi/2}$ displays a $Z_4$ parafermionic signature. In a similar fashion, another set of $\pm \pi/2$ modes near the right end can be explicitly obtained as $\psi_{\pm \pi/2}^R= e^{\mathrm{i}\frac{\pi}{4}}\left(\gamma_{B,N,-1}\mp \gamma_{B,N,1}\gamma_{A,N,-1}\gamma_{B,N,-1} \right)/\sqrt{2}$, which also satisfies $(\psi_{\pm \pi/2}^R)^2=\mathrm{i} \gamma_{B,N,1}\gamma_{A,N,-1}$ and $(\psi_{\pm \pi/2}^R)^4=1$. 

Equation~(\ref{Floq}) respects a global $Z_2$ symmetry and a number of $Z_4$ symmetries, i.e., $\mathcal{Q}_2^\dagger U_T \mathcal{Q}_2 = U_T$ and $\mathcal{Q}_{4,k}^\dagger U_T \mathcal{Q}_{4,k} = U_T$ with $k=1,\cdots,N$,
\begin{eqnarray}
\mathcal{Q}_2 &=& \prod_{j=1}^N \gamma_{A,j,1}\gamma_{B,j,1}\gamma_{A,j,-1}\gamma_{B,j,-1} \;, \nonumber \\
\mathcal{Q}_{4,k} &=& \left(e^{\mathrm{i} \frac{\pi}{4} (1-\mathrm{i} \gamma_{B,k,1} \gamma_{A,k,-1})(1-\prod_{j=1}^N \mathrm{i} \gamma_{B,j,-1} \gamma_{A,j,1}) } \right)  \nonumber \\
&& \times \left(\prod_{j=1}^N \mathrm{i} \gamma_{A,j,1} \gamma_{B,j,1} \right) \;.
\label{symmetries}
\end{eqnarray}
In particular, it can be verified that $\psi_{\pm \pi/2}^L\mathcal{Q}_{4,1}= \mp \mathrm{i} \mathcal{Q}_{4,1} \psi_{\pm \pi/2}^L$. One may then define another set of $\pm \pi/2$ modes as $\tilde{\psi}_{\pi/2}^{R}=\mathcal{Q}_{4,1} \mathcal{Q}_2 \psi_{\pi/2}^{R}$ and $\tilde{\psi}_{-\pi/2}^{R}=\mathcal{Q}_{4,1}^\dagger \mathcal{Q}_2^\dagger \psi_{-\pi/2}^{R}$. While these additional $\pm \pi/2$ modes are not independent of $\psi_{\pm \pi/2}^{R}$, they satisfy the expected $Z_4$ parafermionic algebra with respect to $\psi_{\pm \pi/2}^{L}$, i.e., $\psi_{\pm \pi/2}^L \tilde{\psi}_{\pm \pi/2}^R = \mp \mathrm{i} \tilde{\psi}_{\pm \pi/2}^R \psi_{\pm \pi/2}^L $. As an immediate consequence, the operator $e^{\mathrm{i} \frac{\pi}{4}} \psi_{\pm \pi/2}^L \tilde{\psi}_{\mp \pi/2}^R$, which commutes with $\mathcal{Q}_{4,k}$ and has eigenvalues of $\pm 1, \pm \mathrm{i}$, can be utilized to encode a nonlocal four-dimensional qudit. 

{\it Parameter variation and disorder.} Away from the ideal case considered above, exact analytical treatment is no longer feasible. To investigate the fate of the expected parafermion $\pm \pi/2$ modes at general parameter values, we will thus resort to numerics. To this end, we define a set of spectral functions \cite{parstudies,DTCrel,DTCrel2}
\begin{equation}
s_{\psi,\epsilon} = \mathcal{N}_{\psi,\epsilon} \sum_{n\in \mathcal{X}} \int_{-\delta}^{\delta} S_{\psi,\epsilon}(\varepsilon_n, \eta)   d\eta  \;, 
\end{equation} 
where $\epsilon\geq 0$, $\mathcal{X}$ is a set of some distinct random integers smaller than the the system's Hilbert space dimension, $\delta\ll 1$, $\mathcal{N}_{\psi,\epsilon}=\frac{1}{\sum_{n\in \mathcal{X}}\int_{-\pi/T}^{\pi/T} S_{\psi,\epsilon}{\varepsilon_n,\eta} d\eta}$, and
\begin{eqnarray}
S_{\psi,\epsilon}(\varepsilon_n, \eta) &=& \sum_{k=-\infty}^{\infty} \sum_{\varepsilon_m} |\langle \varepsilon_n |\psi | \varepsilon_m \rangle |^2 \nonumber \\
&& \times \delta(\varepsilon_n-\varepsilon_m -\eta -\epsilon -2\pi k/T) \;.
\end{eqnarray}
Intuitively, $s_{\psi,\epsilon}$ measures the tendency of an operator $\psi$ to be a quasienergy $\epsilon$ excitation satisfying $U_T^\dagger \psi U_T=e^{-\mathrm{i}\epsilon }\psi $. Indeed, if $s_{\psi,\epsilon}=1$, $\psi$ essentially maps any quasienergy eigenstate $|\varepsilon\rangle$ to $|\varepsilon+\epsilon\rangle$. The presence of MZMs, MPMs, and parafermion $\pm \pi/2$ modes is thus signified by the existence of $\psi_0$, $\psi_\pi$, and $\psi_{\pm \pi/2}$ such that $s_{\psi_0,0}$, $s_{\psi_\pi,\pi}$, and $s_{\psi_{\pm \pi/2},\pm \pi/2}$ are respectively close to $1$.

In the system under consideration, we will first use the operator $\psi_{\pi/2}^L$ in the evaluation of the spectral functions. While such operator no longer represents a system's parafermion $\pm \pi/2$ mode (if it exists) at general parameter values, it is expected to have a significant overlap with the actual parafermion $\pm \pi/2$ mode. Therefore, in a regime where parafermion $\pm \pi/2$ modes are present, $s_{\psi_{ \pi/2}^L, \pi/2}$ will remain finite. Evaluating $s_{\psi_{ \pi/2}^L, 0}$ and $s_{\psi_{ \pi/2}^L, \pi}$ will additionally capture the potential presence of zero and $\pi$ modes at some parameter values. 

\begin{center} 
	\begin{figure}
		\includegraphics[scale=0.5]{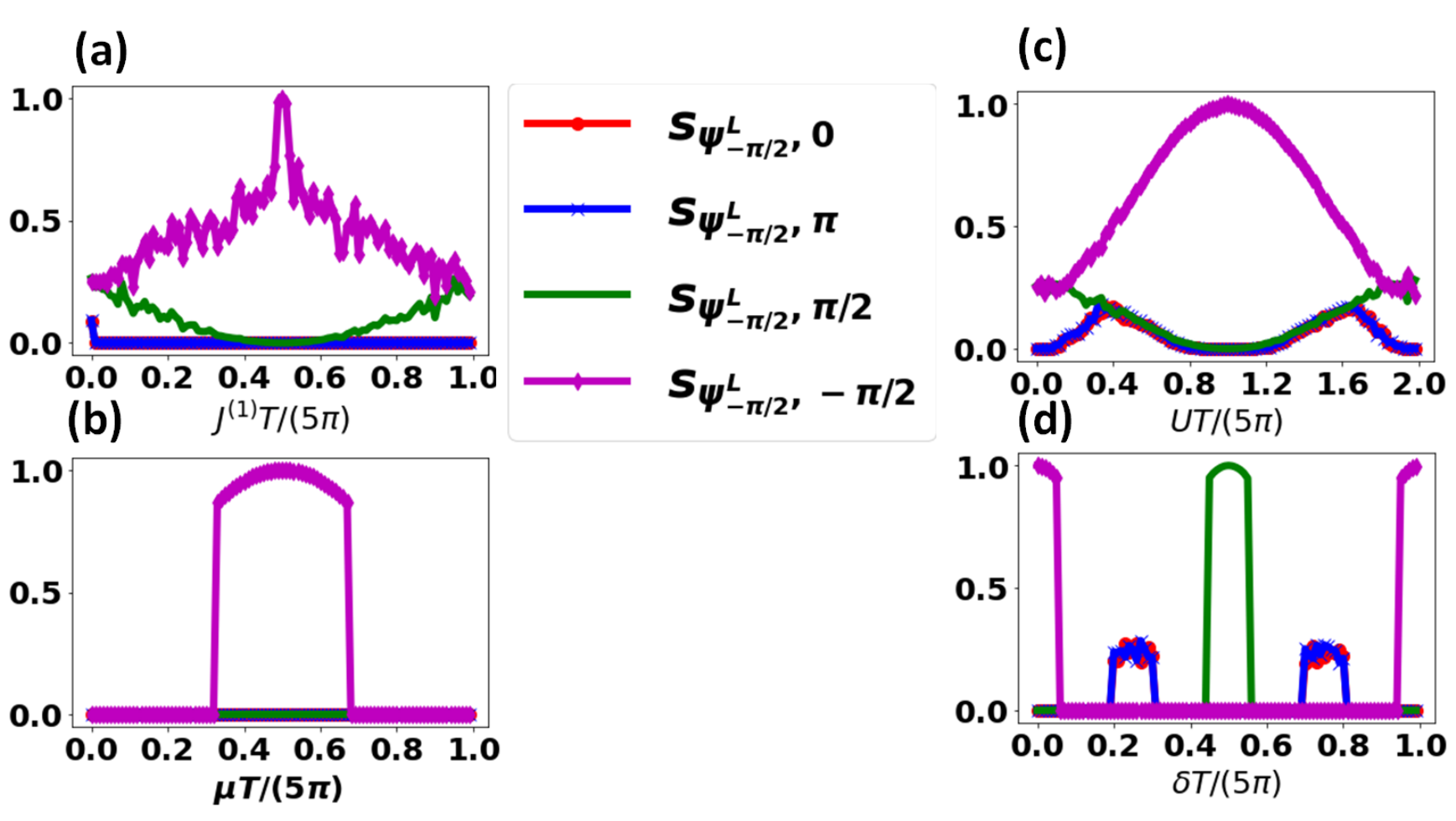}
		\caption{Relevant spectral functions $s_{\psi,\epsilon}$ for probing the presence of zero, $\pi$, and $\pm \pi/2$ modes at varying (a) $J_j^{(1)}=J^{(1)}$, (b) $\mu_{j,s}=\mu$, (c) $U_j=U$, and (d) $\frac{J_{j,s}^{(2)}-\Delta_{j,s}}{2}=\delta$. The fixed parameters are taken as $\frac{J_{j,s}^{(2)}+\Delta_{j,s}}{2}=1.1875\pi/T$, $U_j=2\mu_{j,s}=2J_{2j-1}^{(1)}=5\pi/T$, and $N=4$.}
		\label{fig:intspec}
	\end{figure}
\end{center}

Figure~\ref{fig:intspec} shows the four relevant spectral functions as a function of various system parameters. In particular, it is observed that $s_{\psi_{ -\pi/2}^L, -\pi/2}$ remains finite over a considerable window of parameter values. While not shown in the figure, similar robustness is observed with respect to $s_{\psi_{\pi/2}^L, \pi/2}$, thus highlighting the topological nature of the parafermion $\pm \pi/2$ modes. It is also interesting to note that $s_{\psi_{ -\pi/2}^L, 0}$ and $s_{\psi_{ -\pi/2}^L, \pi}$ become nonzero at a range of parameter values, suggesting the potential presence of zero and $\pi$ modes. As further detailed in \cite{SI}, there is in fact another solvable point at $J_{j,s}^{(2)}-\Delta_{j,s} = \frac{5\pi}{2T}$ where the explicit form of these zero and $\pi$ edge modes, which are fermionic in nature, can be obtained. 

While the quantities $s_{\psi_{ -\pi/2}^L, \epsilon}$ for $\epsilon=0,\pm \pi/2,\pi$ above are useful for detecting the presence of zero, $\pm \pi/2$, and $\pi$ modes, they do not provide information about these modes quasiparticle nature. Indeed, as shown in Fig.~\ref{fig:intspec}(c), $s_{\psi_{ -\pi/2}^L, -\pi/2}$ remains finite even at $U_j=0$. While this agrees with our analysis above that the system still supports $\pm \pi/2$ modes in the noninteracting limit, such $\pm \pi/2$ modes are not parafermionic in nature, i.e., they square to zero as expected from ordinary fermions \cite{SI}. To investigate the fate of parafermion $\pm \pi/2$ modes under varying system parameters, we now evaluate the $\pm \pi/2$ quasienergy spectral functions with respect to the adiabatically deformed $\tilde{\psi}_{\pm \pi/2}^L(s)\equiv \mathcal{U}_T^\dagger(s) \psi_{\pm \pi/2}^L \mathcal{U}_T(s)$, where $\mathcal{U}_T(s)=\mathcal{T}\exp\left(-\mathrm{i} \int_{s_0}^s H_{\rm eff}(s') ds'\right)$, $\mathcal{T}$ is the ordering operator, $s'=J^{(1)},U,\mu,$ or $\delta$ is a slowly changing parameter, and $H_{\rm eff}(s')$ is the effective Hamiltonian which generates the same Floquet operator as Eq.~(\ref{sysraw}). Note in particular that since $\tilde{\psi}_{\pm \pi/2}^L$ represent an almost exact $\pm \pi/2$ mode, provided that it indeed exists and adiabaticity condition holds (no gap closing in the quasienergy excitation spectrum), $s_{\tilde{\psi}_{ \pm \pi/2}^L, \pm \pi/2}$ remains smooth and very close to unity in a regime that is topologically equivalent with the ideal case above. Importantly, since $\mathcal{U}_T(s)$ is unitary, it preserves the algebra of $\tilde{\psi}_{\pm \pi/2}^L(s)$ and the latter remains parafermionic in nature.       

\begin{center} 
	\begin{figure}
		\includegraphics[scale=0.5]{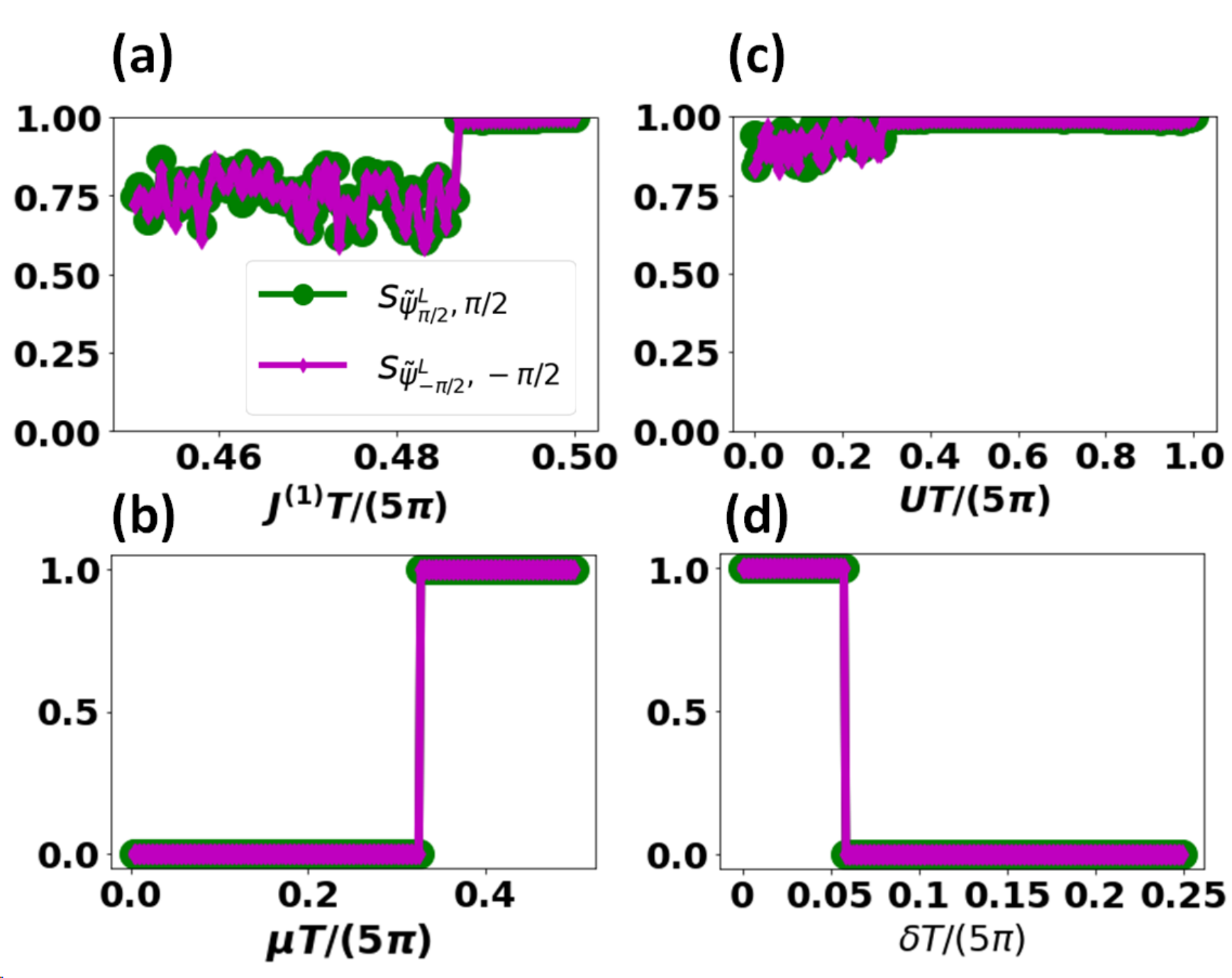}
		\caption{The spectral functions $s_{\tilde{\psi}_{\pm \pi/2},\pm \pi/2}$ with respect to the adiabatically deformed $\psi_{\pm \pi/2}$ at varying (a) $J_j^{(1)}=J^{(1)}$, (b) $\mu_{j,s}=\mu$, (c) $U_j=U$, and (d) $\frac{J_{j,s}^{(2)}-\Delta_{j,s}}{2}=\delta$. The fixed parameters are the same as Fig.~\ref{fig:intspec}.}
		\label{fig:intspec2}
	\end{figure}
\end{center}

Our results are summarized in Fig.~\ref{fig:intspec2}. There, the previously identified topological phase transitions at varying $\mu$ and $\delta$ are now signified by a jump in the associated spectral functions $s_{\tilde{\psi}_{\pm, \pi/2}, \pm \pi/2}$ from $\approx 1$ to $0$. It is worth noting that the clear transition induced by the chemical potential $\mu$ in Fig.~\ref{fig:intspec2}(b) can be exploited to devise a braiding protocol between a pair of $Z_4$ parafermions. To this end, one may reshape the nanowire into a $T$-junction in the spirit of Ref.~\cite{braid1}, then employ a keyboard of locally tunable gates to effectively move the targeted parafermions around each other.

It is also interesting to note that the additional transition point separating the regime supporting parafermionic and ordinary $\pm \pi/2$ modes is now clearly observed in Fig.~\ref{fig:intspec2}(c). Specifically, as $\tilde{\psi}_{\pm \pi/2}$ is obtained from adiabatically deforming the parafermionic $\pm \pi/2$ modes in the ideal case, $s_{\tilde{\psi}_{\pm \pi/2},\pm \pi/2}$ will stay smooth (become noisy while remaining finite) in the regime where such $\pm \pi/2$ modes remain parafermionic in nature (become ordinary fermions). A similar transition is also observed in Fig.~\ref{fig:intspec2}(a). This agrees with our finding, as detailed in \cite{SI}, that the system only supports ordinary fermionic $\pm \pi/2$ modes at $J_j^{(1)}=0$, thus highlighting the importance of Zeeman field in the formation of $Z_4$ parafermions. Additional numerics presented in \cite{SI} further reveals that the fermionic $\pm \pi/2$ modes found in $J_{j}^{(1)}=0$ case and $U_j=0$ case are not topologically equivalent.

\begin{center} 
	\begin{figure}
		\includegraphics[scale=0.35]{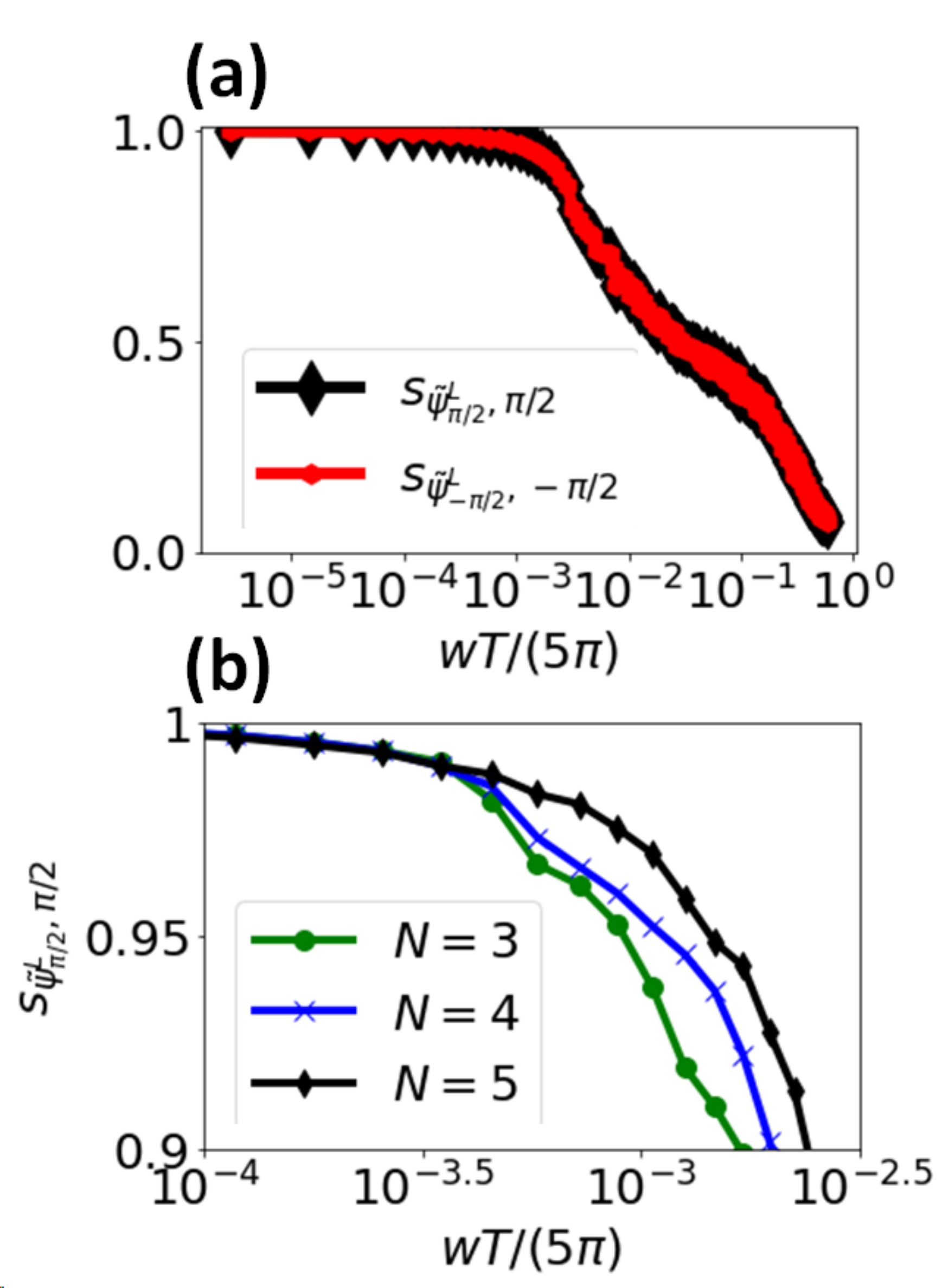}
		\caption{(a) The spectral functions $s_{\tilde{\psi}_{\pm \pi/2},\pm \pi/2}$ at varying disorder strength $w$ and $N=5$. (b) The spectral function $s_{\tilde{\psi}_{\pi/2},\pi/2}$ at three different system sizes. All data points are averaged over $10$ ($50$) disorder realizations for $N=5$ ($N=3$ and $N=4$).}
		\label{fig:intspec3}
	\end{figure}
\end{center}

Finally, we investigate the robustness of the parafermion $\pm \pi/2$ modes against spatial disorder by taking any system parameter $p_{j,s}$ from a uniform distribution $[\bar{p}-\delta p, \bar{p}+\delta p]$. In Fig.~\ref{fig:intspec3}(a), we plot the $\pm \pi/(2T)$ quasienergy spectral functions with respect to the adiabatically deformed $\pm \pi/2$ modes under $2\bar{\mu}=2\bar{J}^{(1)}=4\bar{J}^{(2)}=4\bar{\Delta}=\bar{U}=5\pi/T$ and the the same $\delta \mu=\delta J^{(1)}=\delta J^{(2)}=\delta \Delta =\delta U=w $. Remarkably, a strong signature of $\pm \pi/2$ modes remains present even at disorder strength of $>0.1$, as evidenced by the finite spectral functions. There, the deviation of the spectral functions from the perfect $+1$ value may be attributed to finite size effect that results in the deviation of $\pm \pi/2$ modes from being a true $\pm \pi/(2T)$ quasienergy excitation (Similar to the splitting of Majorana zero modes from zero energy in finite length proximitized nanowires \cite{Majsplit,Majsplit2}). As demonstrated in Fig.~\ref{fig:intspec3}(b), increasing the system size indeed improves the observed spectral functions at moderate disorder.      

{\it Concluding remarks.} We have developed a simple and relatively realistic topological superconducting model capable of hosting parafermion $\pm \pi/2$ edge modes without requiring fractional quantum Hall systems or complicated interactions. This is achieved through the intricate interplay between fermionic Hubbard interaction and appropriately designed periodic driving. Such parafermion $\pm \pi/2$ modes do not only represent examples of topological features with no static and single-particle counterpart, but they may also find a promising application in topological quantum computing \cite{Nayak2008,QEC5}. 

It is worth mentioning that the main idea of our construction, i.e., to design a periodic driving that simulates quantum gate operations on Majorana-based qubits, is very general and can be adapted to construct topological superconductors hosting $\pi/2^n$ parafermion edge modes for any integer $n$ (see \cite{SI}). A simplification of this general construction is expected to be feasible and forms an interesting future work. Devising a protocol for braiding parafermion modes in the proposed setup and its generalization then forms a natural task worth pursuing next. Finally, due to the simplicity of realizing the various terms in our model, experimentally verifying some aspects of this paper in the near future represents another exciting direction. To this end, fractional Josephson effect with $8\pi$ periodicity \cite{par2,parMaj3} and $\pi/(4T)$-bias peak in the two-terminal conductance measurement \cite{Flocond} are some potential experimental signatures of the above $Z_4$ parafermion $\pm \pi/2$ modes to be anticipated.       

\begin{acknowledgements}
	{\bf Acknowledgement}: This work is supported by the Australian Research Council Centre of Excellence for Engineered Quantum Systems (EQUS, CE170100009). 
\end{acknowledgements}

\onecolumngrid 
\appendix

\section{Supplemental Material}

This Supplemental Material consists of three sections. In Section A, we derive Eq.~(4) in the main text from the actual Floquet operator. In Section B, we analytically obtain the zero, $\pm \pi/2$, and $\pi$ modes at four solvable sets parameter values, which correspond to the ideal case, the zero Zeeman field case, the noninteracting case, and the zero superconducting pairing case. Finally, we present a general construction of topological superconducting model supporting parafermion $\pi/2^n$ modes in Section C.

\section{Section A: Obtaining the rotated Floquet operator in the ideal case}
\label{app1}

{Recall that two species of mutually anti-commuting Majorana operators $\gamma_{A,j,\pm 1}$ and $\gamma_{B,j,\pm 1}$ are defined in the main text, such that $c_{j,\pm 1}^\dagger =\frac{\gamma_{B,j,\pm 1} +\mathrm{i} \gamma_{A,j,\pm 1}}{2} $. In terms of these Majorana operators, the five Hamiltonians describing the system over one period can be written as (omitting constant terms)
	
	\begin{eqnarray}
	H_1 &=& \sum_{j=1}^{\frac{N}{2}} \sum_{s=\pm 1} \left( \frac{s\mu}{2} \mathrm{i} \gamma_{A,2j,s} \gamma_{B,2j,s} +\mathrm{i} \frac{J^{(1)}}{2} \gamma_{B,2j-1,s} \gamma_{A,2j-1,-s} \right)  \;, \nonumber \\
	H_2 &=& \sum_{j=1}^{\frac{N}{2}-1} J \left( \mathrm{i} \gamma_{B,2j,1}\gamma_{A,2j-1,1} + \mathrm{i} \gamma_{B,2j+1,-1}\gamma_{A,2j,-1}\right) \;, \nonumber \\
	H_3 &=& \sum_{j=1}^N \frac{u}{4} \left(-\gamma_{A,j,1}\gamma_{B,j,1}\gamma_{A,j,-1}\gamma_{B,j,-1}+\sum_{s=\pm 1} \mathrm{i} \gamma_{A,j,s} \gamma_{B,j,s} \right) \;, \nonumber \\
	H_4 &=& \sum_{j=1}^{\frac{N}{2}-1} J \left(\mathrm{i} \gamma_{B,2j,-1} \gamma_{A,2j-1,-1}+\mathrm{i} \gamma_{B,2j+1,1}\gamma_{A,2j,1}\right) \;, \nonumber \\
	H_5 &=&  \sum_{j=1}^{\frac{N}{2}} \sum_{s=\pm 1} \left( \frac{s\mu}{2} \mathrm{i} \gamma_{A,2j-1,s} \gamma_{B,2j-1,s} -\mathrm{i} \frac{J^{(1)}}{2} \gamma_{B,2j,s} \gamma_{A,2j,-s} \right) \;,
	\end{eqnarray}
	where $u=2\mu=2J^{(1)}=5\pi/T$ in the ideal case. The system's Floquet operator can thus be written explicitly as 
	\begin{eqnarray}
	\tilde{U}_T &=& e^{-\mathrm{i} \frac{H_5 T}{5}} \times e^{-\mathrm{i} \frac{H_4 T}{5}} \times e^{-\mathrm{i} \frac{H_3 T}{5}} \times e^{-\mathrm{i} \frac{H_2 T}{5}} \times e^{-\mathrm{i} \frac{H_1 T}{5}} \nonumber \\ 
	&=& e^{\sum_{j=1}^{\frac{N}{2}} \sum_{s=\pm 1}\frac{\pi}{4} \left(s\gamma_{A,2j-1,s}\gamma_{B,2j-1,s} -\gamma_{B,2j,s} \gamma_{A,2j,-s}\right)} \times e^{\sum_{j=1}^{\frac{N}{2}-1} \frac{JT}{5} \left(\gamma_{B,2j,-1}\gamma_{A,2j-1,-1} + \gamma_{B,2j+1,1}\gamma_{A,2j,1}\right) } \nonumber \\
	&\times&  e^{\sum_{j=1}^N \frac{\pi}{4} \left(\mathrm{i} \gamma_{A,j,1}\gamma_{B,j,1}\gamma_{A,j,-1}\gamma_{B,j,-1}+\sum_{s=\pm 1} \gamma_{A,j,s}\gamma_{B,j,s}\right) } \times e^{\sum_{j=1}^{\frac{N}{2}-1} \frac{JT}{5} \left(\gamma_{B,2j,1}\gamma_{A,2j-1,1} + \gamma_{B,2j+1,-1}\gamma_{A,2j,-1}\right) } \nonumber \\
	&\times& e^{\sum_{j=1}^{\frac{N}{2}} \sum_{s=\pm 1}\frac{\pi}{4} \left(s\gamma_{A,2j,s}\gamma_{B,2j,s} +\gamma_{B,2j-1,s} \gamma_{A,2j-1,-s}\right)} \;, \nonumber \\
	\end{eqnarray}
}
we consider a unitary rotation
\begin{equation}
R=e^{\sum_{j=1}^{\frac{N}{2}} \mathrm{i} \frac{\pi}{4} \gamma_{A,2j-1,1}\gamma_{B,2j-1,1}\gamma_{A,2j-1,-1}\gamma_{B,2j-1,-1}} \times e^{\sum_{j=1}^{\frac{N}{2}} \sum_{s=\pm 1}\frac{\pi}{4}\gamma_{B,2j-1,s}\gamma_{A,2j-1,-s}} \times e^{\sum_{j=1}^{\frac{N}{2}} \frac{\pi}{2}\gamma_{A,2j,1}\gamma_{B,2j,1}}
\end{equation} 
to obtain,
\begin{equation}
R\tilde{U}_TR^\dagger = e^{\sum_{j=1}^N \mathrm{i} \frac{\pi}{4} \gamma_{A,j,1}\gamma_{B,j,1}\gamma_{A,j,-1}\gamma_{B,j,-1} } \times e^{\sum_{j=1}^{N-1} \sum_{s=\pm 1} \frac{\pi}{4} \gamma_{B,j,s}\gamma_{A,j,-s} } \times e^{\sum_{j=1}^{N} \frac{\pi}{2} \gamma_{A,j,1}\gamma_{B,j,1} } \times \Lambda \times \Gamma \;, 
\end{equation} 
where
\begin{eqnarray}
\Lambda &=& e^{-\sum_{j=1}^{\frac{N}{2}} \mathrm{i} \frac{\pi}{4} \gamma_{A,2j,1}\gamma_{B,2j,1}\gamma_{A,2j,-1}\gamma_{B,2j,-1}} \times e^{-\sum_{j=1}^{\frac{N}{2}} \sum_{s=\pm 1}\frac{\pi}{4} \gamma_{A,2j-1,s}\gamma_{B,2j-1,s}} \times e^{\sum_{j=1}^{\frac{N}{2}-1} \frac{JT}{5} \left(\gamma_{B,2j,-1}\gamma_{A,2j-1,-1} + \gamma_{B,2j+1,1}\gamma_{A,2j,1}\right) } \nonumber \\
&\times & e^{\sum_{j=1}^{\frac{N}{2}} \sum_{s=\pm 1}\frac{\pi}{4} \gamma_{A,2j-1,s}\gamma_{B,2j-1,s}} \times e^{\sum_{j=1}^{\frac{N}{2}} \mathrm{i} \frac{\pi}{4} \gamma_{A,2j,1}\gamma_{B,2j,1}\gamma_{A,2j,-1}\gamma_{B,2j,-1}} \;, \nonumber \\
\Gamma &=& e^{\sum_{j=1}^{\frac{N}{2}} \mathrm{i} \frac{\pi}{4} \gamma_{A,2j-1,1}\gamma_{B,2j-1,1}\gamma_{A,2j-1,-1}\gamma_{B,2j-1,-1}} \times e^{\sum_{j=1}^{\frac{N}{2}} \sum_{s=\pm 1}\frac{\pi}{4} \gamma_{A,2j,s}\gamma_{B,2j,s}}  \times e^{\sum_{j=1}^{\frac{N}{2}-1} \frac{JT}{5} \left(\gamma_{B,2j+1,-1}\gamma_{A,2j,-1} + \gamma_{B,2j,1}\gamma_{A,2j-1,1}\right) }  \nonumber \\
&\times & e^{-\sum_{j=1}^{\frac{N}{2}} \sum_{s=\pm 1}\frac{\pi}{4} \gamma_{A,2j,s}\gamma_{B,2j,s}} \times e^{-\sum_{j=1}^{\frac{N}{2}} \mathrm{i} \frac{\pi}{4} \gamma_{A,2j-1,1}\gamma_{B,2j-1,1}\gamma_{A,2j-1,-1}\gamma_{B,2j-1,-1}} \;. \label{Eq:Check} 
\end{eqnarray}
By noting that the argument of each exponential in Eq.~(\ref{Eq:Check}) consists of mutually commuting terms, the latter can be expanded and rearranged. As an example, we may further write
\begin{eqnarray}
\Lambda &=& \prod_{j=1}^{\frac{N}{2}-1} \left(e^{- \mathrm{i} \frac{\pi}{4} \gamma_{A,2j,1}\gamma_{B,2j,1}\gamma_{A,2j,-1}\gamma_{B,2j,-1}} \times e^{-\frac{\pi}{4} \gamma_{A,2j-1,-1}\gamma_{B,2j-1,-1}} \times e^{ \frac{JT}{5} \gamma_{B,2j,-1}\gamma_{A,2j-1,-1} }\right.  \nonumber \\
&\times &\left. e^{\frac{\pi}{4} \gamma_{A,2j-1,-1}\gamma_{B,2j-1,-1}} \times e^{\mathrm{i} \frac{\pi}{4} \gamma_{A,2j,1}\gamma_{B,2j,1}\gamma_{A,2j,-1}\gamma_{B,2j,-1}} \right) \times \left( e^{- \mathrm{i} \frac{\pi}{4} \gamma_{A,2j,1}\gamma_{B,2j,1}\gamma_{A,2j,-1}\gamma_{B,2j,-1}} \times e^{-\frac{\pi}{4} \gamma_{A,2j+1,1}\gamma_{B,2j+1,1}}  \right. \nonumber \\
&\times& \left. e^{ \frac{JT}{5} \gamma_{B,2j+1,1}\gamma_{A,2j,1} } \times e^{\frac{\pi}{4} \gamma_{A,2j+1,1}\gamma_{B,2j+1,1}} \times e^{ \mathrm{i} \frac{\pi}{4} \gamma_{A,2j,1}\gamma_{B,2j,1}\gamma_{A,2j,-1}\gamma_{B,2j,-1}} \right) \;.
\end{eqnarray}
It can further be simplified by utilizing the Euler formula for Pauli-like matrices, i.e.,
\begin{equation}
e^{\mathrm{i} \theta \mathcal{P}_1 } \mathcal{P}_2 e^{-\mathrm{i} \theta \mathcal{P}_1 } = \cos(2\theta) \mathcal{P}_1 +\mathrm{i} \sin(2\theta) \mathcal{P}_1 \mathcal{P}_2\;, \label{Eq:app1}
\end{equation}
where $\mathcal{P}_1^2=\mathcal{P}_2^2=1$ and $\left\lbrace \mathcal{P}_1 ,\mathcal{P}_2 \right\rbrace = 0$. Specifically, 
\begin{eqnarray}
e^{-\frac{\pi}{4} \gamma_{A,2j-1,-1}\gamma_{B,2j-1,-1}} \times e^{ \frac{JT}{5} \gamma_{B,2j,-1}\gamma_{A,2j-1,-1} } \times e^{\frac{\pi}{4} \gamma_{A,2j-1,-1}\gamma_{B,2j-1,-1}} &=& e^{ \frac{JT}{5} \gamma_{B,2j,-1}\gamma_{B,2j-1,-1} } \;, \nonumber \\
e^{- \mathrm{i} \frac{\pi}{4} \gamma_{A,2j,1}\gamma_{B,2j,1}\gamma_{A,2j,-1}\gamma_{B,2j,-1}} \times e^{ \frac{JT}{5} \gamma_{B,2j,-1}\gamma_{B,2j-1,-1} } \times e^{\mathrm{i} \frac{\pi}{4} \gamma_{A,2j,1}\gamma_{B,2j,1}\gamma_{A,2j,-1}\gamma_{B,2j,-1}} &=& e^{ \mathrm{i} \frac{JT}{5} \gamma_{B,2j-1,-1} \gamma_{A,2j,-1} \gamma_{A,2j,1}\gamma_{B,2j,1} } \;, \nonumber \\
\end{eqnarray}
thus simplifying the first round bracket into a single exponential. In a similar fashion, it can be shown that the second round bracket reduces into $e^{\mathrm{i} \frac{JT}{5}\gamma_{B,2j,1} \gamma_{A,2j+1,1}\gamma_{A,2j,-1}\gamma_{B,2j,-1}}$. By repeating the same analysis, $\Gamma$ can be similarly simplified. That is,
\begin{eqnarray}
\Lambda &=& e^{\mathrm{i} \frac{JT}{5} \left(\gamma_{B,2j-1,-1} \gamma_{A,2j,-1} \gamma_{A,2j,1}\gamma_{B,2j,1}+\gamma_{B,2j,1} \gamma_{A,2j+1,1}\gamma_{A,2j,-1}\gamma_{B,2j,-1}\right) } \;, \nonumber \\
\Gamma &=& e^{\mathrm{i} \frac{JT}{5} \left(\gamma_{B,2j,-1} \gamma_{A,2j+1,-1} \gamma_{A,2j+1,1}\gamma_{B,2j+1,1}+\gamma_{B,2j-1,1} \gamma_{A,2j,1}\gamma_{A,2j-1,-1}\gamma_{B,2j-1,-1}\right) } \;,
\end{eqnarray}
and Eq.~(4) in the main text immediately follows.

\section{Section B: Constructing $\pm \pi/2$ modes in the solvable cases}
\label{app2}

\subsection{1. Parafermion $\pm \pi/2$ modes in the ideal case, i.e., $U_j=2\mu_{j,s}=2J_{j}^{(1)}=\frac{5\pi}{T}$ and $J_{j,s}^{(2)}=\Delta_{j,s}=J$ }

Equation.~(4) in the main text can further be decomposed to
\begin{eqnarray}
U_T &=& G_3\times G_2 \times G_1 \;, \nonumber \\
G_3 &=& \prod_{j=1}^N e^{\mathrm{i} \frac{\pi}{4} \gamma_{A,j,1}\gamma_{B,j,1}\gamma_{A,j,-1}\gamma_{B,j,-1}} \times e^{ \frac{\pi}{4} \gamma_{B,j,1}\gamma_{A,j,-1} } \times  e^{ \frac{\pi}{4} \gamma_{B,j,-1}\gamma_{A,j,1} }  \;, \; G_2 = \prod_{j=1}^N e^{\frac{\pi}{2} \gamma_{A,j,1}\gamma_{B,j,1} }  \;, \nonumber \\
G_1 &=& \prod_{j=1}^{N-1} e^{-\mathrm{i} \frac{JT}{5} \gamma_{B,j,-1}\gamma_{A,j+1,-1}\gamma_{A,j+1,1} \gamma_{B,j+1,1}} \times e^{-\sum_{j=1}^{N-1} \mathrm{i} \frac{JT}{5} \gamma_{B,j,1}\gamma_{A,j+1,1}\gamma_{A,j,-1} \gamma_{B,j,-1}} \;. \label{Eq:app2}
\end{eqnarray}
Note that the product of Majorana operators appearing on each exponential of Eq.~(\ref{Eq:app2}) squares to unity and either commutes or anticommutes with $\gamma_{A,1,1}$, thus allowing us to repeatedly apply Eq.~(\ref{Eq:app1}). In particular, $G_1$ fully commutes with $\gamma_{A,1,1}$ and leaves it invariant. There is only one exponential in $G_2$ which anticommutes with $\gamma_{A,1,1}$, so that
\begin{equation}
G_2^\dagger \gamma_{A,1,1} G_2 = e^{-\frac{\pi}{2}\gamma_{A,1,1}\gamma_{B,1,1}} \gamma_{A,1,1} e^{\frac{\pi}{2}\gamma_{A,1,1}\gamma_{B,1,1}} = -\gamma_{A,1,1} \;. 
\end{equation}
Finally,
\begin{eqnarray}
-G_3^\dagger \gamma_{A,1,1} G_3 &=& - e^{-\mathrm{i} \frac{\pi}{4} \gamma_{A,1,1}\gamma_{B,1,1}\gamma_{A,1,-1}\gamma_{B,1,-1}}  e^{ -\frac{\pi}{4} \gamma_{B,1,-1}\gamma_{A,1,1} } \gamma_{A,1,1} e^{ \frac{\pi}{4} \gamma_{B,1,-1}\gamma_{A,1,1} } e^{\mathrm{i} \frac{\pi}{4} \gamma_{A,1,1}\gamma_{B,1,1}\gamma_{A,1,-1}\gamma_{B,1,-1}}  \nonumber \\
&=& e^{-\mathrm{i} \frac{\pi}{4} \gamma_{A,1,1}\gamma_{B,1,1}\gamma_{A,1,-1}\gamma_{B,1,-1}}\gamma_{B,1,-1} e^{\mathrm{i} \frac{\pi}{4} \gamma_{A,1,1}\gamma_{B,1,1}\gamma_{A,1,-1}\gamma_{B,1,-1}} \nonumber \\
&=& -\mathrm{i} \gamma_{A,1,1}\gamma_{B,1,1}\gamma_{A,1,-1} \;.
\end{eqnarray}
This shows that $U_T^\dagger \gamma_{A,1,1} U_T =-\mathrm{i} \gamma_{A,1,1}\gamma_{B,1,1}\gamma_{A,1,-1}$, as claimed in the main text. In a similar fashion, it is easily seen that $-\mathrm{i} \gamma_{A,1,1}\gamma_{B,1,1}\gamma_{A,1,-1}$ commutes with $G_1$ and $G_2$. On the other hand,
\begin{eqnarray}
-\mathrm{i} G_3^\dagger \gamma_{A,1,1}\gamma_{B,1,1}\gamma_{A,1,-1} G_3 &=& -\mathrm{i} e^{-\mathrm{i} \frac{\pi}{4} \gamma_{A,1,1}\gamma_{B,1,1}\gamma_{A,1,-1}\gamma_{B,1,-1}}  e^{ -\frac{\pi}{4} \gamma_{B,1,-1}\gamma_{A,1,1} } \gamma_{A,1,1}\gamma_{B,1,1}\gamma_{A,1,-1} e^{ \frac{\pi}{4} \gamma_{B,1,-1}\gamma_{A,1,1} } \nonumber \\
&& \times e^{\mathrm{i} \frac{\pi}{4} \gamma_{A,1,1}\gamma_{B,1,1}\gamma_{A,1,-1}\gamma_{B,1,-1}}  \nonumber \\
&=& \mathrm{i} e^{-\mathrm{i} \frac{\pi}{4} \gamma_{A,1,1}\gamma_{B,1,1}\gamma_{A,1,-1}\gamma_{B,1,-1}}\gamma_{B,1,1}\gamma_{A,1,-1}\gamma_{B,1,-1} e^{\mathrm{i} \frac{\pi}{4} \gamma_{A,1,1}\gamma_{B,1,1}\gamma_{A,1,-1}\gamma_{B,1,-1}} \nonumber \\
&=& -\gamma_{A,1,1} \;.
\end{eqnarray}
By defining $\gamma = a \gamma_{A,1,1} -b \mathrm{i} \gamma_{A,1,1}\gamma_{B,1,1}\gamma_{A,1,-1}$, the eigenvalue equation $U_T^\dagger \gamma U_T = e^{-\mathrm{i} \epsilon} \gamma$ can be equivalently recast in matrix form as
\begin{equation}
\left(\begin{array}{cc}
0 & -1 \\
1 & 0
\end{array} \right) \left(\begin{array}{c}
a  \\
b 
\end{array} \right) =  e^{-\mathrm{i} \epsilon}\left(\begin{array}{c}
a  \\
b 
\end{array} \right) \;.
\end{equation}
Such a matrix has eigenvalues of $\pm \mathrm{i}$, which correspond to $\epsilon=\mp \pi/2$. The associated eigenvectors, i.e., $(a,b)= (1,\pm \mathrm{i})$ then enable the construction of $\pm \pi/2$ modes presented in the main text. 

\subsection{2. Ordinary fermion $\pm \pi/2$ modes in the zero Zeeman field case, i.e., $J_j^{(1)}=0$}

\begin{center} 
	\begin{figure}
		\includegraphics[scale=1]{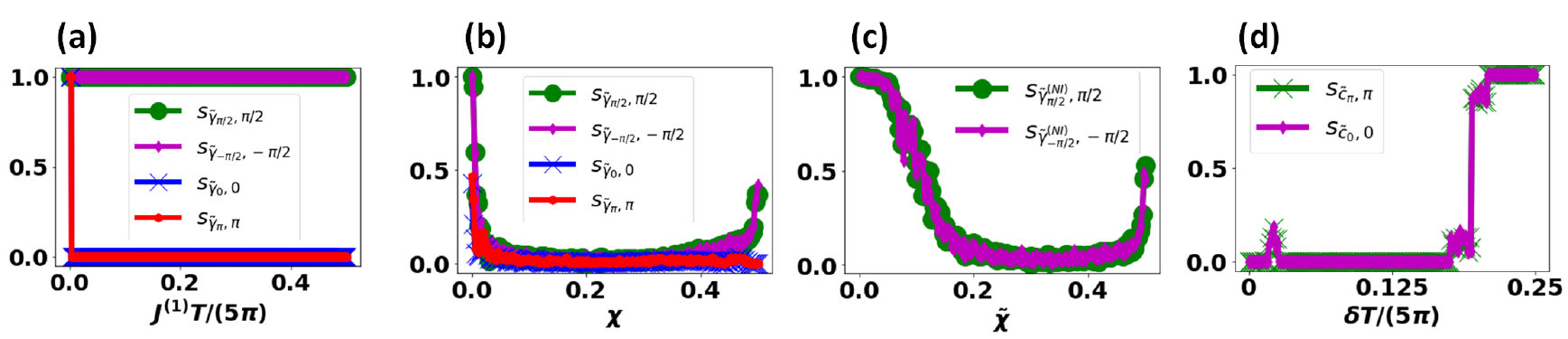}
		\caption{(a,b) The relevant spectral functions evaluated with respect to the adiabatic evolution of $\gamma_0$, $\gamma_{\pm \pi/2}$, and $\gamma_\pi$ in Eq.~(\ref{ordmodes}) at varying (a) $J_j^{(1)}=J^{(1)}$ and (b) $2J_j^{(1)}T/(5\pi)=1-U_jT/(5\pi)=2\chi $. (c) The relevant spectral functions evaluated with respect to the adiabatic evolution of $\gamma_{\pm \pi/2}^{\rm (NI)}$ in Eq.~(\ref{ordmodes2}) at varying $1-2J_j^{(1)}T/(5\pi)=U_jT/(5\pi)=2\tilde{\chi}$. (d) The relevant spectral functions evaluated with respect to the adiabatic of $c_{0}$ and $c_{\pi}$ in Eq.~(\ref{ordmodes3}) at varying $\frac{J_{j,s}^{(2)}-\Delta_{j,s}}{2}=\delta$. In all panels, the fixed parameters are taken as $\frac{J_{j,s}^{(2)}+\Delta_{j,s}}{2}=1.1875\pi/T$, $U_j=2\mu_{j,s}=2J_{2j-1}^{(1)}=5\pi/T$, and $N=4$.}
		\label{fig:intspec4}
	\end{figure}
\end{center}

For simplicity, we further take $U_j=2\mu_{j,2}=4J_{j,s}^{(2)}=4\Delta_{j,s}=\frac{5\pi}{T}$. It is easier to analyze the actual (non-rotated) Floquet operator directly, which can be written as
\begin{eqnarray}
\tilde{U}_T &=& G_5\times G_4 \times G_3 \times G_2 \times G_1 \;, \nonumber \\
G_5 &=& \prod_{j=1}^{\frac{N}{2}} \prod_{s=\pm 1} e^{\frac{s\pi}{4}\gamma_{A,2j-1,s} \gamma_{B,2j-1,s}} \;, \; G_4 = \prod_{j=1}^{\frac{N}{2}-1} e^{\frac{\pi}{4} \gamma_{B,2j,-1} \gamma_{A,2j-1,-1} } \times e^{\frac{\pi}{4} \gamma_{B,2j+1,1} \gamma_{A,2j,1} } \;, \nonumber \\
G_3 &=& \prod_{j=1}^N e^{\mathrm{i} \frac{\pi}{4} \gamma_{A,j,1}\gamma_{B,j,1}\gamma_{A,j,-1}\gamma_{B,j,-1}}  \times e^{ \frac{\pi}{4} \gamma_{A,j,1}\gamma_{B,j,1} } \times e^{ \frac{\pi}{4} \gamma_{A,j,-1}\gamma_{B,j,-1} }  \;, \nonumber \\
G_2 &=& \prod_{j=1}^{\frac{N}{2}-1} e^{\frac{\pi}{4} \gamma_{B,2j,1}\gamma_{A,2j-1,1} } \times e^{\frac{\pi}{4} \gamma_{B,2j+1,-1}\gamma_{A,2j,-1} }  \;, \; G_1 = \prod_{j=1}^{\frac{N}{2}} \prod_{s=\pm 1} e^{\frac{s\pi}{4} \gamma_{A,2j,s}\gamma_{B,2j,s}} \;. \label{Eq:app3}
\end{eqnarray}
Using Eq.~(\ref{Eq:app1}), we may now explicitly evaluate $\tilde{U}_T^\dagger \gamma_{A,1,1} \tilde{U}_T$. To this end, by denoting $\xrightarrow{U}$ as $U^\dagger \cdots U$, we obtain
\begin{eqnarray}
\gamma_{A,1,1} &\xrightarrow{G_1}& \gamma_{A,1,1} \xrightarrow{G_2} -\gamma_{B,2,1} \xrightarrow{G_3} \mathrm{i} \gamma_{B,2,1} \gamma_{A,2,-1} \gamma_{B,2,-1} \xrightarrow{G_4} \mathrm{i} \gamma_{B,2,1} \gamma_{A,2,-1} \gamma_{A,1,-1} \nonumber \\
&\xrightarrow{G_5}&  -\mathrm{i} \gamma_{B,2,1} \gamma_{A,2,-1} \gamma_{B,1,-1} \;, \nonumber \\
-\mathrm{i} \gamma_{B,2,1} \gamma_{A,2,-1} \gamma_{B,1,-1} &\xrightarrow{G_1}& -\mathrm{i} \gamma_{A,2,1} \gamma_{B,2,-1} \gamma_{B,1,-1} \xrightarrow{G_2} -\mathrm{i} \gamma_{A,2,1} \gamma_{B,2,-1} \gamma_{B,1,-1}   \nonumber \\
& \xrightarrow{G_3}& \gamma_{B,2,1} \gamma_{A,2,-1} \gamma_{A,1,1} \gamma_{B,1,1} \gamma_{B,1,-1} \xrightarrow{G_4} \gamma_{B,2,1} \gamma_{A,2,-1} \gamma_{A,1,1} \gamma_{B,1,1} \gamma_{B,1,-1}  \nonumber \\
&\xrightarrow{G_5}& \gamma_{B,2,1} \gamma_{A,2,-1} \gamma_{A,1,1} \gamma_{B,1,1} \gamma_{A,1,-1} \;, \nonumber \\
\gamma_{B,2,1} \gamma_{A,2,-1} \gamma_{A,1,1} \gamma_{B,1,1} \gamma_{A,1,-1} &\xrightarrow{G_1}& \gamma_{A,2,1} \gamma_{B,2,-1} \gamma_{A,1,1} \gamma_{B,1,1} \gamma_{A,1,-1} \xrightarrow{G_2} -\gamma_{A,2,1} \gamma_{B,2,-1} \gamma_{B,2,1} \gamma_{B,1,1} \gamma_{A,1,-1} \nonumber \\
&\xrightarrow{G_3} & -\mathrm{i} \gamma_{B,2,-1} \gamma_{A,1,1} \gamma_{B,1,-1} \xrightarrow{G_4} -\mathrm{i} \gamma_{A,1,-1} \gamma_{A,1,1} \gamma_{B,1,-1} \xrightarrow{G_5} -\mathrm{i} \gamma_{A,1,-1} \gamma_{B,1,1} \gamma_{B,1,-1} \;, \nonumber \\
-\mathrm{i} \gamma_{A,1,-1} \gamma_{B,1,1} \gamma_{B,1,-1} &\xrightarrow{G_1}& -\mathrm{i} \gamma_{A,1,-1} \gamma_{B,1,1} \gamma_{B,1,-1} \xrightarrow{G_2} -\mathrm{i} \gamma_{A,1,-1} \gamma_{B,1,1} \gamma_{B,1,-1} \xrightarrow{G_3} -\gamma_{B,1,1} \xrightarrow{G_4} -\gamma_{B,1,1} \nonumber \\
&\xrightarrow{G_5}& \gamma_{A,1,1}\;.  
\end{eqnarray}
We may then use the ansatz 
\begin{equation}
\gamma = A\;\gamma_{A,1,1} -B\; \mathrm{i} \gamma_{B,2,1}\gamma_{A,2,-1}\gamma_{B,1,-1} + C \; \gamma_{B,2,1}\gamma_{A,2,-1}\gamma_{A,1,1}\gamma_{B,1,1} \gamma_{A,1,-1} -D\; \mathrm{i} \gamma_{A,1,-1} \gamma_{B,1,1} \gamma_{B,1,-1} \;,    
\end{equation}
so that the eigenvalue equation $\tilde{U}_T^\dagger \gamma \tilde{U}_T=e^{-\mathrm{i} \epsilon} \gamma$ becomes the matrix equation
\begin{equation}
\left(\begin{array}{cccc}
0 & 1 & 0 & 0 \\
0 & 0 & 1 & 0 \\
0 & 0 & 0 & 1 \\
1 & 0 & 0 & 0
\end{array} \right) \left(\begin{array}{c}
A  \\
B \\
C \\
D \\
\end{array} \right) =  e^{-\mathrm{i} \epsilon}\left(\begin{array}{c}
A  \\
B \\
C \\
D 
\end{array} \right) \;.
\end{equation}
It is easily verified that such a matrix has eigenvalues of $\pm 1$ and $\pm \mathrm{i}$, thus corresponding to $\epsilon=0,\pm\frac{\pi}{2},\pi$. The associated zero, $\pm \pi/2$, and $\pi$ modes can then be written as
\begin{eqnarray}
\gamma_{0}&=& \frac{1}{2} \left(\gamma_{A,1,1} - \mathrm{i} \gamma_{B,2,1}\gamma_{A,2,-1}\gamma_{B,1,-1} +  \gamma_{B,2,1}\gamma_{A,2,-1}\gamma_{A,1,1}\gamma_{B,1,1} \gamma_{A,1,-1} - \mathrm{i} \gamma_{A,1,-1} \gamma_{B,1,1} \gamma_{B,1,-1} \right) \;, \nonumber \\
\gamma_{\pi}&=& \frac{1}{2} \left(\gamma_{A,1,1} + \mathrm{i} \gamma_{B,2,1}\gamma_{A,2,-1}\gamma_{B,1,-1} +  \gamma_{B,2,1}\gamma_{A,2,-1}\gamma_{A,1,1}\gamma_{B,1,1} \gamma_{A,1,-1} + \mathrm{i} \gamma_{A,1,-1} \gamma_{B,1,1} \gamma_{B,1,-1} \right) \;, \nonumber \\
\gamma_{\pm \pi/2}&=& \frac{1}{2} \left(\pm\gamma_{A,1,1} -  \gamma_{B,2,1}\gamma_{A,2,-1}\gamma_{B,1,-1} \mp \gamma_{B,2,1}\gamma_{A,2,-1}\gamma_{A,1,1}\gamma_{B,1,1} \gamma_{A,1,-1} + \gamma_{A,1,-1} \gamma_{B,1,1} \gamma_{B,1,-1} \right) \;. \label{ordmodes}
\end{eqnarray}
Note however that $\gamma_{\pm \pi/2}^2=0$, which implies that these $\pm \pi/2$ modes are merely ordinary fermions. In addition, as shown in Fig.~\ref{fig:intspec4}(a), the above zero and $\pi$ modes quickly disappear as some system parameters are tuned away from the specific values above. 

\subsection{3. Ordinary fermion $\pm \pi/2$ modes in the noninteracting limit}

By further taking $\mu_{j,s}=J_j^{(1)}=\frac{5\pi}{2T}$ and $J_{j,s}^{(2)}=\Delta_{j,s}=J$, we obtain
\begin{eqnarray}
\tilde{U}_T &=& G_3 \times G_2 \times G_1 \;, \nonumber \\
G_3 &=& \prod_{j=1}^{\frac{N}{2}} \prod_{s=\pm 1} e^{\frac{s\pi}{4}\gamma_{A,2j-1,s} \gamma_{B,2j-1,s}} \times e^{-\frac{\pi}{4}\gamma_{B,2j,s} \gamma_{A,2j,-s}} \;, \; G_2 = \prod_{j=1}^{N-1} \prod_{s=\pm 1} e^{\frac{JT}{5} \gamma_{B,j+1,s}\gamma_{A,j,s} }  \;, \nonumber \\
G_1 &=& \prod_{j=1}^{\frac{N}{2}} \prod_{s=\pm 1} e^{\frac{s\pi}{4}\gamma_{A,2j,s} \gamma_{B,2j,s}} \times e^{\frac{\pi}{4}\gamma_{B,2j-1,s} \gamma_{A,2j-1,-s}} \;. \label{Eq:app4}
\end{eqnarray}
We may then evaluate
\begin{eqnarray}
\gamma_{A,1,1} &\xrightarrow{G_1}& -\gamma_{B,1,-1} \xrightarrow{G_2} -\gamma_{B,1,-1} \xrightarrow{G_3} -\gamma_{A,1,-1} \;,  \nonumber \\
-\gamma_{A,1,-1} &\xrightarrow{G_1}& \gamma_{B,1,1} \xrightarrow{G_2} \gamma_{B,1,1} \xrightarrow{G_3}  -\gamma_{A,1,1} \;.
\end{eqnarray}
Repeating the same procedure as before, we identify 
\begin{equation}
\gamma_{\pm \pi/2}^{(\rm NI)}= \gamma_{A,1,1}\pm \mathrm{i} \gamma_{A,1,-1} \label{ordmodes2} 
\end{equation}
as the $\pm \pi/2$ modes. However, as $\left(\gamma_{\pm \pi/2}^{{(\rm NI)}}\right)^2=0$, it follows that these $\pm \pi/2$ modes correspond to ordinary fermions. 

In Fig.~\ref{fig:intspec4}(b,c), we plot the spectral functions associated with the adiabatically evolved $\pm \pi/2$ modes from the zero Zeeman field case of Section~B2 to the noninteracting case and vice versa. It is found that lowering the interaction strength quickly kills the $\pm \pi/2$ modes from the zero Zeeman field case. On the other hand, the $\pm \pi/2$ modes from the noninteracting case are found to survive up to a moderate interaction strength before they disappear and reemerge at the parameter values of Section~B2. These results demonstrate that, despite both $\pm \pi/2$ modes being ordinary fermions, they are not topologically equivalent.

\subsection{4. Ordinary fermion zero and $\pi$ modes at $J_{j,s}^{(2)}-\Delta_{j,s}=\frac{5\pi}{2T}$}

For further simplicity, we take $J_{j,s}^{(2)}=\mu_{j,s}=J_j^{(1)}=\frac{5\pi}{2T}$ and $\Delta_{j,s}=0$. The Floquet operator can then be written as
\begin{eqnarray}
\tilde{U}_T &=& G_5\times G_4 \times G_3 \times G_2 \times G_1 \;, \nonumber \\
G_5 &=& \prod_{j=1}^{\frac{N}{2}} \prod_{s=\pm 1} e^{\frac{s\pi}{4}\gamma_{A,2j-1,s} \gamma_{B,2j-1,s}} \times e^{\frac{\pi}{4} \gamma_{A,2j,s} \gamma_{B,2j,-s}} \;, \nonumber \\
G_4 &=& \prod_{j=1}^{\frac{N}{2}-1} e^{\frac{\pi}{4} \gamma_{B,2j,-1} \gamma_{A,2j-1,-1} } \times e^{\frac{\pi}{4} \gamma_{B,2j-1,-1} \gamma_{A,2j,-1} } \times e^{\frac{\pi}{4} \gamma_{B,2j+1,1} \gamma_{A,2j,1} } \times e^{\frac{\pi}{4} \gamma_{B,2j,1} \gamma_{A,2j+1,1} } \;, \nonumber \\
G_3 &=& \prod_{j=1}^N e^{\mathrm{i} \frac{\pi}{4} \gamma_{A,j,1}\gamma_{B,j,1}\gamma_{A,j,-1}\gamma_{B,j,-1}} \times e^{ \frac{\pi}{4} \gamma_{A,j,1}\gamma_{B,j,1} } \times  e^{ \frac{\pi}{4} \gamma_{A,j,-1}\gamma_{B,j,-1} }  \;, \nonumber \\
G_2 &=& \prod_{j=1}^{\frac{N}{2}-1} e^{\frac{\pi}{4} \gamma_{B,2j,1}\gamma_{A,2j-1,1} } \times e^{\frac{\pi}{4} \gamma_{B,2j-1,1}\gamma_{A,2j,1} } \times  e^{\frac{\pi}{4} \gamma_{B,2j+1,-1}\gamma_{A,2j,-1} } \times e^{\frac{\pi}{4} \gamma_{B,2j,-1}\gamma_{A,2j+1,-1} }  \;,\nonumber \\
G_1 &=& \prod_{j=1}^{\frac{N}{2}} \prod_{s=\pm 1} e^{\frac{s\pi}{4} \gamma_{A,2j,s}\gamma_{B,2j,s}} \times e^{\frac{\pi}{4} \gamma_{B,2j-1,s} \gamma_{A,2j-1,-s}} \;. \label{Eq:app5}
\end{eqnarray}
Following the same steps as before, we evaluate the conjugation of $\gamma_{A,1,1}$ by $\tilde{U}_T$ as
\begin{eqnarray}
\gamma_{A,1,1} &\xrightarrow{G_1}& -\gamma_{B,1,-1} \xrightarrow{G_2} -\gamma_{B,1,-1} \xrightarrow{G_3} \mathrm{i} \gamma_{A,1,1} \gamma_{B,1,1} \gamma_{B,1,-1} \xrightarrow{G_4} \mathrm{i} \gamma_{A,1,1} \gamma_{B,1,1} \gamma_{A,2,-1} \xrightarrow{G_5} \mathrm{i} \gamma_{A,1,1} \gamma_{B,1,1} \gamma_{B,2,1} \;, \nonumber \\
\mathrm{i} \gamma_{A,1,1} \gamma_{B,1,1} \gamma_{B,2,1} &\xrightarrow{G_1} & \mathrm{i} \gamma_{B,1,-1} \gamma_{A,1,-1} \gamma_{A,2,1} \xrightarrow{G_2} -\mathrm{i} \gamma_{B,1,-1} \gamma_{A,1,-1} \gamma_{B,1,1} \xrightarrow{G_3} -\gamma_{B,1,1} \xrightarrow{G_4} -\gamma_{B,1,1} \xrightarrow{G_5} \gamma_{A,1,1} \;. 
\end{eqnarray}
This implies that $\gamma^\alpha_0=\gamma_{A,1,1}+\mathrm{i} \gamma_{A,1,1} \gamma_{B,1,1} \gamma_{B,2,1}$ and $\gamma^\alpha_\pi=\gamma_{A,1,1}-\mathrm{i} \gamma_{A,1,1} \gamma_{B,1,1} \gamma_{B,2,1}$ are zero and $\pi$ modes. However, these modes are unphysical since $(\gamma^\alpha_0)^2\propto \gamma^{\alpha}_0$ and $(\gamma^\alpha_\pi)^2\propto \gamma^{\alpha}_\pi$. It turns out that another set of left-localized zero and $\pi$ modes exist, which can be found by conjugating $\gamma_{A,2,1}$ with $\tilde{U}_T$. That is,
\begin{eqnarray}
\gamma_{A,2,1} &\xrightarrow{G_1}& \gamma_{B,2,1} \xrightarrow{G_2} \gamma_{A,1,1} \xrightarrow{G_3} -\mathrm{i} \gamma_{A,1,1} \gamma_{A,1,-1} \gamma_{B,1,-1} \xrightarrow{G_4} \mathrm{i} \gamma_{A,1,1} \gamma_{B,2,-1} \gamma_{A,2,-1} \xrightarrow{G_5} -\mathrm{i} \gamma_{B,1,1} \gamma_{A,2,1} \gamma_{B,2,1} \;, \nonumber \\
-\mathrm{i} \gamma_{B,1,1} \gamma_{A,2,1} \gamma_{B,2,1} &\xrightarrow{G_1} & -\mathrm{i} \gamma_{A,1,-1} \gamma_{A,2,1} \gamma_{B,2,1} \xrightarrow{G_2} \mathrm{i} \gamma_{A,1,-1} \gamma_{B,1,1} \gamma_{A,1,1} \xrightarrow{G_3} \gamma_{A,1,-1} \xrightarrow{G_4} -\gamma_{B,2,-1} \xrightarrow{G_5} \gamma_{A,2,1} \;,
\end{eqnarray}
so that $\gamma^\beta_0=\gamma_{A,2,1}-\mathrm{i} \gamma_{B,1,1} \gamma_{A,2,1} \gamma_{B,2,1}$ and $\gamma^\beta_\pi=\gamma_{A,2,1}+\mathrm{i} \gamma_{B,1,1} \gamma_{A,2,1} \gamma_{B,2,1}$ are another set of zero and $\pi$ modes. In particular, the superpositions
\begin{eqnarray}
c_{0}&=& \gamma_0^\alpha +\mathrm{i} \gamma_0^\beta \;, \nonumber \\
c_\pi &=& \gamma_\pi^\alpha +\mathrm{i} \gamma_\pi^\beta \;. \label{ordmodes3}
\end{eqnarray}
are also zero and $\pi$ modes. Since $c_0^2=c_\pi^2=0$, they represent ordinary fermions. In Fig.~\ref{fig:intspec4}(d), we adiabatically evolve $c_{0}$ and $c_{\pi}$ as the parameter $J_{j,s}^{(2)}-\Delta_{j,s}$ is varied and plot the relevant spectral functions. It is observed that the regime in which $c_{0}$ and $c_{\pi}$ exist agree with the prediction of Fig.~2 in the main text.



\section{Section C: Recipe for constructing interacting periodically driven topological superconductors supporting parafermion $\pi/2^n$ modes}
\label{app3}

Consider an array of $n+1$ spinless superconducting chains, each of which is labelled as $\nu=1,\cdots, n+1$ and of size $N$. We will now show that appropriate interaction among the different chains, together with systematic periodic modulation of system parameters, allows the emergence of $\pi/2^n$ parafermion modes at the edges of the superconducting chains. 

The main idea of our construction is based on the unitary
\begin{eqnarray}
\tilde{U}_T^{(2^n)} &=& \left(\prod_{k=2}^{n+1} e^{-\mathrm{i} \sum_{j=1}^N \frac{\pi}{2^k} \left[\prod_{\nu=1}^{k-1} (1-Z_{j,\nu} Z_{j,\nu+1}))\right] (1-\prod_{\nu=1}^k X_{j,\nu}) } \right) \times e^{-\sum_{j=1}^N \mathrm{i} \frac{\pi}{2} X_{j,1}}  \nonumber \\
&& \times e^{- \mathrm{i} \sum_{j=1}^{N-1} \left(\sum_{\nu=1}^n J_{j,\nu} Z_{j,\nu}Z_{j+1,\nu}Z_{j,\nu+1}Z_{j+1,\nu+1} +J_{j,n+1} Z_{j,n+1} Z_{j+1,n+1} \right)  }\;, \label{genuni} 
\end{eqnarray}
where $X_{j,\nu}$, $Y_{j,\nu}$, and $Z_{j,\nu}$ are Pauli operators associated with a spin living on site $(j,\nu)$ of a $N\times (n+1)$ rectangular lattice. Specifically, Eq.~(\ref{genuni}) can be mapped to
\begin{equation}
U_T^{(2^n)} = \left( \prod_{k=2}^{n+1} e^{-\mathrm{i} \sum_{j=1}^N \frac{\pi}{2^k} \left[\prod_{\nu=1}^{k-1} (1-Z_{j,\nu}) \right] (1-X_{j,k}) } \right) \times e^{-\sum_{j=1}^N \mathrm{i} \frac{\pi}{2} X_{j,1}}  \times e^{-\mathrm{i} \sum_{\nu=1}^{n+1} \sum_{j=1}^{N-1} J_{j,\nu}  Z_{j,\nu} Z_{j+1,\nu} }\;, \label{genuni2}  
\end{equation}
via the unitary transformation $U_T^{(2n)} = u \tilde{U}_T^{(2n)} u^\dagger$, where $u = \prod_{k=n}^{1} e^{-\sum_{j=1}^N \mathrm{i} \frac{\pi}{4}(1-Z_{j,k+1}) X_{j,k} }$. Equation~(\ref{genuni2}) can be analytically diagonalized \cite{repDTC}, the quasienergies of which can be grouped into $n$-tuplets with $\frac{\pi}{2^n T}$ spacing. Since a unitary transformation preserves the system's quasienergies, it follows that quasienergies of $\tilde{U}_T^{(2n)}$ also form $n$-tuplets with $\frac{\pi}{2^n T}$ spacing. 

The eigenstates of $U_T^{(2^n)}$ can be explicitly obtained as
\begin{eqnarray}
|\varepsilon_{\left\lbrace \mathcal{J}_{j,\nu} \right\rbrace ,\ell}\rangle &=& \sum_{j=0}^{2^n-1} e^{\mathrm{i} \frac{j\pi \ell}{2^{n-1}} } |\bar{m} \rangle_{\left\lbrace \mathcal{J}_{j,\nu} \right\rbrace}\;,   
\end{eqnarray}
where $|\bar{m} \rangle_{\left\lbrace \mathcal{J}_{j,\nu} \right\rbrace} \equiv |\bar{s}_1\cdots \bar{s}_n\rangle_{\left\lbrace \mathcal{J}_{j,\nu} \right\rbrace} $ such that $\mathcal{J}_{j,\nu},\bar{s}_k=\pm 1$, $\bar{m}=\left(\sum_{k=0}^{n} \bar{s}_{k+1} 2^k \right) \;\text{mod}\;2$, $Z_{j,\nu}Z_{j+1,\nu}|\bar{m}\rangle_{\mathcal{J}_{j,\nu}} =\mathcal{J}_{j,\nu} |\bar{m}\rangle_{\mathcal{J}_{j,\nu}}$, and $Z_{a,k}|\bar{m}\rangle_{\left\lbrace \mathcal{J}_{j,\nu} \right\rbrace} = \bar{s}_{k}|\bar{m}\rangle_{\left\lbrace \mathcal{J}_{j,\nu} \right\rbrace}$. An imperfect application of $U_T^{(2^n)}$ can always be written as $U_{T,\rm imperfect}^{(2^n)}=U'U_T^{(2^n)}$, where $U'$ can be expanded as a superposition of products of Pauli operators $Z_{j,\nu}$ and $X_{j,\nu}$ \cite{repDTC}. Using Floquet perturbation theory, the lowest order quasienergy correction to $\varepsilon_{\left\lbrace \mathcal{J}_{j,\nu} \right\rbrace ,\ell}$ can be written as $\mathrm{i}\;\mathrm{log}\left(\langle \varepsilon_{\left\lbrace \mathcal{J}_{j,\nu} \right\rbrace ,\ell} | U' |\varepsilon_{\left\lbrace \mathcal{J}_{j,\nu} \right\rbrace ,\ell} \rangle\right)$. Since $|\varepsilon_{\mathcal{J}_{j,\nu},\ell}\rangle $ only consists of states belonging to the $\left\lbrace \mathcal{J}_{j,\nu}\right\rbrace$ subspace, any strings of low-weight $X_{j,\nu}$ operators will transform $|\varepsilon_{\mathcal{J}_{j,\nu},\ell}\rangle $ to an orthogonal state outside this subspace. On the other hand, while a weight-one $Z_{j,\nu}$ Pauli operator preserves the stabilizer subspace, it can be easily checked that $\langle \varepsilon_{\left\lbrace \mathcal{J}_{j,\nu} \right\rbrace ,\ell} | Z_{a,\alpha} |\varepsilon_{\left\lbrace \mathcal{J}_{j,\nu} \right\rbrace ,\ell} \rangle=0$. Therefore, the nonzero contributions to the quasienergy correction $\mathrm{i}\;\mathrm{log}\left(\langle \varepsilon_{\left\lbrace \mathcal{J}_{j,\nu} \right\rbrace ,\ell} | U' |\varepsilon_{\left\lbrace \mathcal{J}_{j,\nu} \right\rbrace ,\ell} \rangle\right)$ only come from terms in $U'$ that are $\propto \prod_{a=1}^N X_{a\alpha}$. Since such terms only appear at the order of at least $N$ in some imperfection parameter $\epsilon$, the quasienergy structure of $\tilde{U}_T^{(2n)}$, hence the resulting $\frac{\pi}{2^nT}$ quasienergy excitations, becomes more robust with increasing system size.

Next, we apply a generalized Jordan-Wigner transformation to map the above spin-1/2 rectangular lattice into $n+1$ chains of size-$N$ $1D$ Majorana lattices under the Floquet operator
\begin{eqnarray}
\tilde{U}_{T,\rm Maj}^{(2^n)} &=& \left(\prod_{k=2}^{n+1} e^{-\mathrm{i} \sum_{j=1}^N \frac{\pi}{2^k} \left[\prod_{\nu=1}^{k-1} (1-\mathrm{i} \gamma_{B,j,\nu} \gamma_{A,j,\nu+1}))\right] \left[1-\prod_{\nu=1}^k (\mathrm{i} \gamma_{A,j,\nu}\gamma_{B,j,\nu} )\right] } \right) \times e^{-\mathrm{i} \frac{\pi}{2} \sum_{j=1}^N \left(\mathrm{i} \gamma_{A,j,1} \gamma_{B,j,1}\right)} \nonumber \\
&& \times e^{\mathrm{i} \sum_{j=1}^{N-1} \left(\sum_{\nu=1}^{n} J_{j,\nu}  \gamma_{B,j,\nu} \gamma_{A,j,\nu+1} \gamma_{B,j+1,\nu} \gamma_{A,j+1,\nu+1} - \mathrm{i} J_{j,n+1} \gamma_{B,j,n+1} \left(\prod_{\nu=1}^n \mathrm{i} \gamma_{A,j+1,\nu} \gamma_{B,j+1,\nu} \right) \gamma_{A,j+1,n+1} \right)} \;, \label{maj2n}
\end{eqnarray}
where 
\begin{eqnarray}
\gamma_{A,j,\nu} &=& \left( \prod_{k<j} \prod_{\alpha=1}^{n+1} X_{k,\alpha}\right) \left(\prod_{\beta<\nu} X_{j,\beta} \right) Z_{j,\nu} \;, \nonumber \\
\gamma_{B,j,\nu} &=& \left( \prod_{k<j} \prod_{\alpha=1}^{n+1} X_{k,\alpha}\right) \left(\prod_{\beta<\nu} X_{j,\beta} \right) Y_{j,\nu} \;.
\end{eqnarray}
In this case, the presence of $\frac{\pi}{2^n T}$ quasienergy excitation identified above leads to the emergence of $Z_{2^n}$ parafermion modes at the ends of each chain. Finally, a corresponding superconducting model can be obtained by further replacing the Majorana operators appearing in Eq.~(\ref{maj2n}) with complex fermions $c_{j,\nu}=\gamma_{B,j,\nu}-\mathrm{i} \gamma_{A,j,\nu}$.

It should be noted that while the above construction always yields a physical model, i.e. due to the even number of Majorana operators appearing in all terms of Eq.~(\ref{maj2n}), it may not be optimal; the corresponding superconducting model may require long range and intricate interaction that may be challenging to realize in experiments. However, significant simplification can typically be made on a case-by-case basis, e.g., through appropriate unitary transformation. In addition, the fine tuning of various parameters to $\frac{\pi}{2^k}$ in Eq.~(\ref{maj2n}) may not be necessary in its actual implementation due to its inherent quantum error correction mechanism \cite{repDTC}.   

\subsection{1. Application to parafermion $\pm \pi/2$ modes}

To demonstrate the above construction at work, we apply it to obtain a superconducting model supporting parafermion $\pm \pi/2$ modes from two chains of spinless $p$-wave superconductors (or alternatively, a chain of spinful $p$-wave superconductor), subjected to the Floquet operator 
\begin{eqnarray}
\tilde{U}_T^{(2)} &=& e^{-\sum_{j=1}^N \mathrm{i} \frac{\pi}{4} (1-Z_{j,1}Z_{j,2}) (1-X_{j,1}X_{j,2})} \times e^{-\sum_{j=1}^N \mathrm{i} \frac{\pi}{2} X_{j,1}} \times e^{-\mathrm{i} \sum_{j=1}^{N-1} \left(J_{j,1} Z_{j,1}Z_{j+1,1}Z_{j,2}Z_{j+1,2} + J_{j,2} Z_{j,2}Z_{j+1,2}\right) } \nonumber \\
&\rightarrow & e^{-  \sum_{j=1}^N  \frac{\pi}{4} \left(\mathrm{i} \gamma_{A,j,1}\gamma_{B,j,1}\gamma_{A,j,2}\gamma_{B,j,2} + \sum_{s=\pm 1} \gamma_{B,j,(3+s)/2}\gamma_{A,j,(3-s)/2} \right) } \times e^{\sum_{j=1}^{N} \frac{\pi}{2} \gamma_{A,j,1}\gamma_{B,j,1} } \nonumber \\
&& \times e^{\mathrm{i} \sum_{j=1}^{N-1}  \left(J_{j,1} \gamma_{B,j,1}\gamma_{A,j,2}\gamma_{B,j+1,1} \gamma_{A,j+1,2} + J_{j,2} \gamma_{B,j,2}\gamma_{A,j+1,1}\gamma_{B,j+1,1} \gamma_{A,j+1,2} \right) }   \;. \label{ex2}
\end{eqnarray}
Apart from the last exponential, it is noted that Eq.~(\ref{ex2}) is equivalent to Eq.~(4) considered in the main text. Nevertheless, directly rewriting all Majorana operators of Eq.~(\ref{ex2}) in terms of complex fermions yields an effective superconducting model
\begin{eqnarray}
\tilde{U}_T^{(2)} &=& e^{-\mathrm{i} H_3} \times e^{-\mathrm{i} H_2} \times e^{-\mathrm{i} H_1} \;, \nonumber \\
H_1 &=& \sum_{j=1}^{N-1} \left\lbrace J_{j,1} \left(-c_{j,1}^\dagger c_{j,2}+c_{j,1}^\dagger c_{j,2}^\dagger +h.c. \right) \left(-c_{j+1,1}^\dagger c_{j+1,2}+c_{j+1,1}^\dagger c_{j+1,2}^\dagger +h.c.\right) \right. \nonumber \\ 
&+&  2J_{j,2} \left( c_{j,2}^\dagger c_{j+1,2}\hat{n}_{j+1,1} - c_{j,2}^\dagger c_{j+1,2}^\dagger \hat{n}_{j+1,1} \right) +\left. J_{j,2} \left( - c_{j,s}^\dagger c_{j+1,2}  + c_{j,2}^\dagger c_{j+1,2}^\dagger\right)  +h.c.\right\rbrace    \nonumber \\
H_2 &=& \sum_{j=1}^N \frac{\pi}{2} c_{j,1}^\dagger c_{j,1} \;, \nonumber \\
H_3 &=& \sum_{j=1}^{N} \left\lbrace \pi \hat{n}_{j,1}\hat{n}_{j,2}- \frac{\pi}{2} c_{j,1}^\dagger c_{j,2} +h.c. \right\rbrace  -\sum_{j=1}^{N} \sum_{s=1,2} \frac{\pi}{2} c_{j,s}^\dagger c_{j,s} \;. 
\end{eqnarray}  
Note that the periodically driven Hamiltonian realizing the above Floquet operator is much more complex as compared with Eq.~(1) presented in the main text; not only does it involve more system parameters, but it also demands complicated interaction beyond the modest fermion Hubbard type. It is interesting to note that interaction of the form $H_1$ has also been proposed to obtain parafermion zero modes in static superconducting chains \cite{parMaj,parMaj2}. While such intricate interaction seems to be truly necessary in static systems, the system presented in the main text demonstrates the possibility of utilizing periodic driving to simplify the type of interaction required for generating parafermion modes.

\end{document}